\documentclass[a4paper, 11pt, oneside]{article}
\pdfoutput=1
\usepackage{aapreprint}
\makeatletter
\def\@fpheader{\href{http://inspirehep.net/record/1635289}{\textcolor{gray}{JHEP\,{\bf02}\,(2018)\,048}}}
\makeatother

\usepackage{graphicx,wrapfig,float,slashed}
\usepackage{amsmath,amssymb,dsfont,epsfig,graphicx}
\usepackage{epstopdf}
\usepackage{ragged2e}
\usepackage[utf8]{inputenc}
\usepackage{epstopdf}
\usepackage[usenames,dvipsnames,table]{xcolor}
\usepackage{tabu}
\usepackage[toc]{appendix}
\usepackage[normalem]{ulem}
\usepackage[font={small,it},skip=0pt]{caption}
\allowdisplaybreaks
\usepackage{pifont}
\usepackage{subcaption}
\usepackage{microtype}

\newcommand{\nc}{\newcommand}
\nc{\non}{\nonumber}
\nc{\hc}{\hbox {H.c.}}
\nc{\noi}{\noindent}
\nc{\barx}{\bar{x}}
\nc{\pbarn}{\;\hbox {pb}}
\nc{\fbarn}{\;\hbox {fb}}

\nc{\hsp}{\hspace{0.5cm}}
\nc{\lsp}{\hspace{1cm}}
\nc{\Lsp}{\hspace{2cm}}
\nc{\LLsp}{\lsp\lsp}
\nc{\lra}{\longrightarrow}
\nc{\p}{\prime}
\nc{\sgn}{\text{sgn}}
\nc{\ph}{\varphi}
\nc{\op}{{\cal O}}
\nc{\cL}{{\cal L}}
\nc{\tr}{{\text{Tr}}}
\nc{\eq}{\text{Eq.~}}
\nc{\cg}{{\cal G}}
\nc{\ch}{{\cal H}}
\nc{\cZ}{\mathbb Z}
\nc{\cw}{\cos\theta_{\textsc w}}
\nc{\sw}{\sin\theta_{\textsc w}}
\nc{\cwsq}{\cos^2\theta_{\textsc w}}
\nc{\swsq}{\sin^2\theta_{\textsc w}}
\nc{\beq}{\begin{equation}}  \nc{\eeq}{\end{equation}}
\nc{\bea}{\begin{eqnarray}}  \nc{\eea}{\end{eqnarray}}
\nc{\baa}{\begin{array}}     \nc{\eaa}{\end{array}}
\nc{\bit}{\begin{itemize}}   \nc{\eit}{\end{itemize}}
\nc{\ben}{\begin{enumerate}} \nc{\een}{\end{enumerate}}
\nc{\bce}{\begin{center}}    \nc{\ece}{\end{center}}
\nc{\bpm}{\begin{pmatrix}}   \nc{\epm}{\end{pmatrix}}
\nc{\bvt}{\begin{verbatim}}  \nc{\evt}{\end{verbatim}}
\def\lsim{\mathrel{\raise.3ex\hbox{$<$\kern-.75em\lower1ex\hbox{$\sim$}}}}
\def\gsim{\mathrel{\raise.3ex\hbox{$>$\kern-.75em\lower1ex\hbox{$\sim$}}}}

\def\udots{\mathinner{\mkern1mu\raise1pt\vbox{\kern7pt\hbox{.}}\mkern2mu\raise4pt\hbox{.}\mkern2mu\raise7pt\hbox{.}\mkern1mu}}

\def\gev{\;\hbox{GeV}}
\def\tev{\;\hbox{TeV}}

\definecolor{agray}{rgb}{0.95, 0.95, 0.99}
\def\mht{m_{\hat h}}
\def\mhsm{m_{h}}
\def\htw{\hat{h}}

\def\ie{{\it i.e.}}
\def\eg{{\it e.g.}}

\def\mth{\textsc{mth}}
\def\fth{\textsc{fth}}
\def\eff{\text{eff}}

\newcommand\fverb{\setbox\fverbbox=\hbox\bgroup\verb}
\newcommand\fverbdo{\egroup\medskip\noindent%
			\fbox{\unhbox\fverbbox}\ }
\newcommand\fverbit{\egroup\item[\fbox{\unhbox\fverbbox}]}
\newbox\fverbbox

\title{Heavy Higgs of the Twin Higgs Models}
\author{Aqeel Ahmed}
\affiliation{Faculty of Physics, University of Warsaw, Pasteura 5, 02-093 Warsaw, Poland\\
PRISMA Cluster of Excellence \& MITP, Johannes Gutenberg University, 55099 Mainz, Germany\\
Laboratoire de Physique Th\'eorique, CNRS, Universit\'e Paris-Sud 11, F-91405 Orsay Cedex, France
}
\emailAdd{aqeel.ahmed@fuw.edu.pl}

\abstract{
Twin Higgs models are the prime illustration of neutral naturalness, where the new particles of the twin sector, gauge singlets of the Standard Model (SM), ameliorate the little hierarchy problem. In this work, we analyse phenomenological implications of the heavy Higgs of the Mirror Twin Higgs and Fraternal Twin Higgs models, when electroweak symmetry breaking is linearly realized. The most general structure of twin Higgs symmetry breaking, including explicit soft and hard breaking terms in the scalar potential, is employed. The direct and indirect searches at the LHC are used to probe the parameter space of Twin Higgs models through mixing of the heavy Higgs with the SM Higgs and decays of the heavy Higgs to the SM states. Moreover,  for the Fraternal Twin Higgs, we study the production and decays of twin glueball and bottomonium states to the SM light fermions, which have interesting signatures involving displaced vertices and are potentially observable at the colliders.}

\preprint{MITP/17--080}
\arxivnumber{1711.03107}

\begin{document}

\maketitle
\flushbottom
\section{Introduction}
\label{Introduction}
The discovery of a scalar boson with mass 125 GeV at the Large~Hadron~Collider~(LHC) and its properties show that it is consistent with the Higgs boson of the Standard Model~(SM). The SM of particle physics is the most accurate theory of elementary particles and their interactions, however, it is not a complete theory of nature since there are many unanswered puzzles which go beyond the Standard Model (BSM). In particular, the discovery of a relatively light SM Higgs has highlighted the need to address {\it naturalness/hierarchy}  problem more than ever before, \ie\ there should be some mechanism which cancels the SM radiative corrections to the Higgs mass. Motivated by naturalness, enormous work has been done in the last few decades constructing BSM scenarios. Most of the BSM models including supersymmetry, composite Higgs, and extra dimensions predict some sort of new physics (top partners) just above the electroweak scale to cancel the quantum corrections to the Higgs mass. However, the absence of such new physics signatures at the LHC pose a ``little hierarchy'' (fine-tuning) problem. Hence, there is a need to construct and analyse new BSM scenarios which can explain the elusiveness of the top-partners and solve the SM puzzles. 

{\it Neutral Naturalness} is one such class of models, where the top-partners are not charged under the SM gauge symmetry, so they are very hard to produce at the colliders. Twin Higgs (TH) model \cite{Chacko:2005pe,Barbieri:2005ri,Chacko:2005vw,Chacko:2005un} is the prime and first example of neutral naturalness, which is designed to resolve the little hierarchy problem and make the top-partners elusive. In a nutshell, TH models have two core ingredients: (i) the SM-like Higgs is a Goldstone boson which emerges due to spontaneous breaking of a global symmetry, hence its mass is protected, and (ii) the twin top partners are neutral (``colorless'') under the SM gauge group, which explains their elusiveness at the LHC. In twin Higgs mechanism the cancellations of radiative corrections to the SM Higgs mass are due to the exchange of new twin states which are related to the SM by a discrete $\cZ_2$ symmetry. 

The simplest realization of twin Higgs mechanism~\cite{Chacko:2005pe} employs an $SU(4)$ global symmetry which is spontaneously broken to $SU(3)$ at a high energy scale $\Lambda$ and gives {\bf 7}~Goldstone bosons and a radial mode. A subgroup $SU(2)\!\times\!\widehat{SU}(2)$ of $SU(4)$ global symmetry is gauged, where $SU(2)$ and $\widehat{SU}(2)$ represent the SM and twin weak gauge groups, respectively ({\it hat} ``~$\widehat{~}$~'' denotes twin sector). After TH electroweak symmetry breaking (EWSB), ${\bf 6}$~of the Goldstone bosons of $SU(4)/SU(3)$ coset are ``eaten'' by the weak gauge bosons of the SM $(W^\pm,\!Z)$ and twin $(\hat W^\pm,\!\hat Z)$ sectors. The remaining one Goldstone boson serves as the SM Higgs. The original incarnation of this idea, where the twin sector has an exact mirror copy of the SM particle content, is referred as `Mirror Twin Higgs' (MTH) model~\cite{Chacko:2005pe}. 

However, in order to manifest the features of the twin Higgs mechanism, one does not need to have an exact copy of the SM particle content in the twin sector, but rather a minimal number of states are necessary to ensure the cancellation of radiative corrections to the SM Higgs mass. This possibility was first discussed in the so-called `Fraternal Twin Higgs' (FTH) model~\cite{Craig:2015pha}. In order to solve the little hierarchy problem, the FTH model requires only the third generation of quarks and leptons, and the weak gauge bosons in the twin sector. Another striking difference of the FTH model to the MTH case is twin hadron spectroscopy in the two models. In particular, the FTH model has two interesting features: ({\em i}) unlike the MTH model, there are no fraternal twin hadrons with light quark flavors ($\hat u,\hat d,\hat s, \hat c$), but the fraternal twin hadrons are glueball~$[\hat g\hat g]$ and bottomonium~$[\hat b\hat b]$ states~\cite{Craig:2015pha}, and ({\em ii}) the fraternal twin QCD confinement scale, $\hat \Lambda_{\textsc{qcd}}$, can be much larger than that of the MTH model, which is due to fewer twin quark flavors and faster running of the twin QCD couplings~$\hat \alpha_{\textsc{qcd}}$. Hence, the  fraternal twin hadrons (glueballs/bottomoniums) can be relatively heavier~\cite{Craig:2015pha}. Note that the twin hadrons are colorless under the SM QCD. The lightest stable twin hadrons in the MTH model are the mirror baryons which could be dark matter candidates~\cite{Barbieri:2005ri}, however, in the FTH case the stable twin hadronic states are the lightest glueball/bottomonium states~\cite{Craig:2015pha}. Moreover, among the fraternal twin hadrons, the $0^{++}$ glueball and bottomonium states have the correct quantum numbers to mix with the scalar sector of the model\,\footnote{The $0^{++}$ twin hadrons are also present in the MTH model, however, given the usual spectroscopy of mirror twin hadrons, these states have low production multiplicity and are unstable (similar to SM $0^{++}$ hadrons). Hence, phenomenologically they have less impact than the fraternal $0^{++}$ twin hadrons.}. This mixing leads to exotic decays of the fraternal $0^{++}$ twin hadrons back into the SM light quarks and leptons at the colliders~\cite{Craig:2015pha}. These features of the FTH model are similar to those of the `hidden valley' models \cite{Strassler:2006im,Strassler:2006ri,Han:2007ae,Kang:2008ea,Juknevich:2009ji,Juknevich:2009gg}. 

There are many variants of TH models which offer UV complete description of the TH mechanism; including supersymmetric TH~\cite{Burdman:2006tz,Falkowski:2006qq,Chang:2006ra,Craig:2013fga,Katz:2016wtw,Badziak:2017syq,Badziak:2017kjk}, composite/holographic TH~\cite{Batra:2008jy,Craig:2014aea,Geller:2014kta,Barbieri:2015lqa,Low:2015nqa,Csaki:2015gfd}, 
and some more recent TH constructions \cite{Craig:2014roa,Burdman:2014zta,Beauchesne:2015lva,Yu:2016bku,Craig:2016kue,Harnik:2016koz,Barbieri:2016zxn,Chacko:2016hvu,Barbieri:2017opf,Serra:2017poj,Csaki:2017jby}. In most of the TH models, it is assumed that the radial mode of TH EWSB is of the order of symmetry breaking scale or heavier. Hence the radial mode is integrated out and low-energy effective theory contains only the Goldstone bosons. However, if the TH spontaneous symmetry breaking is linearly realized then, in principle, the radial mode can be much lighter than the symmetry breaking scale and can be present in the low-energy theory\,\footnote{A possibility of twin Higgs radial mode is also discussed in~\cite{Barbieri:2005ri,Craig:2015pha,Buttazzo:2015bka,Katz:2016wtw}. Moreover, see~\cite{Fichet:2016xvs,Fichet:2016xpw} for the phenomenology of the radial mode in composite Higgs model.}, in fact, is favored by the electroweak precisions tests (EWPT) (see below, Sec.~\ref{ewpt}). Note that the TH symmetry breaking can be linearly realized in a perturbative UV complete TH model, e.g. supersymmetric TH~\cite{Katz:2016wtw}. However, we work in an effective field theory (EFT) framework of the TH models without considering any specific perturbative UV complete scenario. In this work, we explore the scalar sector of the TH mechanism with a focus on the phenomenological implications of the radial mode. The twin radial mode is heavier than the SM-like Higgs of mass $125\gev$, hence we refer it as `heavy' twin Higgs.  

Twin Higgs scalar sector provides a portal between the visible (SM) and invisible (twin) sectors. Hence, the discovery potential of twin Higgs mechanism needs a detailed phenomenological analysis of the scalar sector of TH models. Recently, there has been some work done on different aspects of the TH scalar sector involving only the SM Higgs, see \eg\ \cite{Craig:2015pha,Craig:2014roa,Burdman:2014zta,Beauchesne:2015lva,Yu:2016bku,Craig:2016kue,Harnik:2016koz,Barbieri:2016zxn,Chacko:2016hvu,Barbieri:2017opf,Curtin:2015fna,Csaki:2015fba,Chacko:2015fbc,Pierce:2017taw}. In this work, we aim to investigate the phenomenological aspects of the scalar sector of MTH and FTH models focusing on heavy twin Higgs. We parametrize the fundamental scalar of $SU(4)$, which breaks the global symmetry, by two Higgs doublets $H$ and $\widehat H$. We consider the most general form of TH scalar potential with soft and hard $\mathbb{Z}_2$ discrete and $SU(4)$ global symmetry breaking terms. The effective TH scalar potential has five parameters, however, after fixing the SM-like Higgs mass $m_h\!=\!125\gev$ and its vacuum expectation value (vev) $v\!=\!246\gev$, we are left with three free parameters; the heavy twin Higgs mass, its vev, and the hard $\mathbb{Z}_2$ breaking parameter. In this work, we explore the parameter space spanned by these three free parameters of the TH models at the LHC. Note that relatively small number of free parameters offers a unique opportunity to discover the twin Higgs mechanism at the current and/or future colliders. 

To analyse phenomenological aspects of the scalar sector of the MTH and FTH models, we calculate the branching fractions of the heavy Higgs to the SM and dark (twin) states. It turns out that dominant decay channels of heavy twin Higgs are pairs of the SM Higgs, SM weak gauge bosons, and twin weak gauge bosons (if kinematically allowed), roughly with equal branching fractions. This makes the heavy Higgs decays to the SM massive gauge bosons and the SM Higgs as the prime channels to discover the TH models at the colliders. We perform a detailed analysis of these channels at the LHC with $14\tev$ center-of-mass energy. Moreover, indirect searches at the LHC, via the SM Higgs signal-strength measurements and its invisible decay width, provide a complementary probe to the scalar sector of the TH models. 

\begin{wrapfigure}{r}{0.5\textwidth}
\centering
\includegraphics[width=0.5\textwidth]{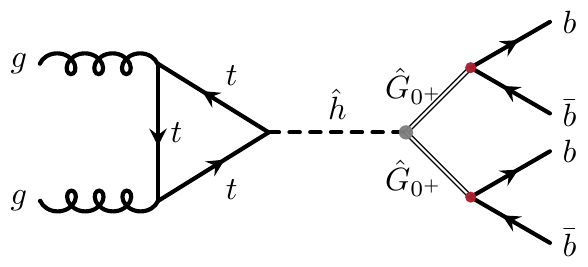}
\caption{An illustration of a benchmark process of $0^{++}$ twin glueball $\hat G_{0^+}$ production via heavy twin Higgs $\hat h$ at the LHC and decays to the SM bottom quarks through an off-shell SM Higgs.}
\label{fig:twingbproddecay}
\end{wrapfigure}

In the case of FTH, the phenomenological study of the twin hadrons (twin glueball and bottomonium states) is much more involved, mainly due to the lack of proper knowledge of twin QCD confinement/hadronization. However, for our purposes, we make conservative estimates of the twin hadron production through the heavy twin Higgs. The production rates for the twin $0^{++}$ hadrons through heavy twin Higgs are about the same order as those of the SM Higgs and massive gauge bosons. These twin $0^{++}$~hadrons (glueball and/or bottomonium states) decay back to the SM light quarks and leptons through the Higgs mixing, see a prototype process in Fig.~\ref{fig:twingbproddecay}. The $0^{++}$ twin hadrons decays are either prompt or displaced depending on their kinematics. Especially, the displaced signals offer a unique discovery potential for the FTH model at the colliders. \hfill~
%

The paper is organized as follows. Twin Higgs symmetry breaking details, the choice of physical basis, and EWPT constraints on TH models are presented in Sec.~\ref{Linear twin Higgs models}. The phenomenological analysis of the twin Higgs of the MTH and FTH models are presented in Secs.~\ref{Heavy Higgs Phenomenology in the Mirror Twin Higgs} and \ref{Heavy Higgs Phenomenology in the Fraternal Twin Higgs}, respectively. Section~\ref{Conclusions} contains the summary and conclusions. We collect supplementary material in Appendix~\ref{Partial decay widths of heavy twin Higgs}, which includes Feynman rules and partial decay widths of the heavy twin Higgs to the SM and twin sector states.

\section{Electroweak Symmetry Breaking in Twin Higgs Models}
\label{Linear twin Higgs models}
We consider the TH models where an $SU(4)$ global symmetry is spontaneously broken to $SU(3)$ providing {\bf 7}~Goldstone bosons. For the TH symmetry breaking we employ a scalar $\mathbb{H}$ which belongs to fundamental representation of $SU(4)$ and parametrize as, 
\beq
\mathbb{H}=\frac1{\sqrt2}\exp\left(\frac{i\Pi}{f_0} \right)\bpm0\\0\\0\\ \ch\epm,     \label{H_su4}
\eeq
where $\Pi$ is a matrix containing the {\bf 7}~Goldstone bosons and $\ch$ is a radial Higgs which acquires a vev $\langle\ch\rangle\!\equiv\! f_0$. Twin Higgs potential for the EWSB is 
\beq
V(\mathbb{H})=\lambda\big(\mathbb{H}^\dag \mathbb{H}- f_0^2/2\big)^2\,,        \label{VH}
\eeq
where using~\eqref{H_su4}, we get a quartic potential for the radial Higgs $\ch$, 
\beq
V(\ch)=\frac14\lambda\big(\ch^2-f_0^2\big)^2\,.        \label{VcH}
\eeq
In order to have a global minimum of the above potential we require,  
\beq
\frac{dV(\ch)}{d\ch}\bigg\vert_{\ch\!=\!f_0}\!=0~, \lsp {\rm and}\lsp \frac{d^2V(\ch)}{d\ch^2}\bigg\vert_{\ch\!=\!f_0}\!>0~. 
\eeq
The above conditions insure that mass squared of the radial fluctuation, $m_{\hat h_0}^2\!=\!2\lambda f_0^2$, is positive for $f_0^2\!>\!0$ and $\lambda>0$. Note that mass of the radial mode can be much lower than the TH symmetry breaking scale $f_0$ for $\lambda<1$. Whereas for $\lambda\gg1$, we recover a non-linear sigma model version of the twin Higgs models where the radial mode is decoupled. 

In this work, we parametrize the scalar $\mathbb{H}$ in a linear sigma model framework, \ie
\beq
\mathbb{H}=\bpm H\\\widehat H\epm,    \lsp H\equiv \frac1{\sqrt2}\bpm h_1+ih_2\\ v+h_0+ih_3\epm,    \lsp \widehat H\equiv \frac1{\sqrt2}\bpm  \hat h_1+i\hat h_2\\ \hat v +\hat h_0+i\hat h_3\epm,    \label{H_AB}
\eeq
where $H$ and $\widehat H$ are doublets under the SM and twin $SU(2)$ weak groups, respectively, whereas $v$ and $\hat v$ are their respective vevs. 
The most general TH effective scalar potential including explicit soft and hard $\mathbb{Z}_2$ discrete and $SU(4)$ global symmetry breaking terms can be written as~\cite{Craig:2015pha,Katz:2016wtw},
\begin{align}
V_{\eff}(H,\widehat H)&=\lambda\Big(|H|^2+|\widehat H|^2-\frac{f_0^2}2\Big)^2+\kappa\big(|H|^4+|\widehat H|^4\big)-\sigma f_0^2 |H|^2+\rho |H|^4.        \label{Veff_AB}
\end{align}
Note that one can rewrite all other possible $\mathbb{Z}_2/SU(4)$ explicit symmetry breaking terms in the form of above potential parameters. As mentioned above, we do not aim to consider any specific UV complete description of TH models, rather we assume that the above form of effective potential can arise from a particular UV complete model, \eg~\cite{Katz:2016wtw}.  

Before moving forward, we would like to discuss the role/significance of each term in the above effective potential. 
\ben\itemsep0em
\item The $\lambda$-term is invariant under $SU(4)$ global symmetry as well as respects $\mathbb{Z}_2$ discrete symmetry under which $H\leftrightarrow\widehat H$. As noted above, for $f_0^2\!>\!0$, this terms is the source of spontaneous breaking of $SU(4)$ global symmetry to $SU(3)$ giving {\bf 7}~Goldstone bosons plus a radial mode $\hat h_0$.
\item The $\kappa$-term is invariant under the discrete $\cZ_2$ symmetry, however, breaks $SU(4)$ global symmetry explicitly. For $\kappa\!>\!0$, the vevs $v$ and $\hat v$ are degenerate in the absence of $\cZ_2$ breaking term, \ie\ $v\!=\!\hat v\!=\!f_0/\!\sqrt{\!(2\lambda+\kappa)}$, which is phenomenologically not a viable option.  
\item The $\sigma$-term `softly' breaks both $SU(4)$ global and $\cZ_2$ discrete symmetries, as a result a misalignment of the two vevs is generated. 
\item The $\rho$-term captures `hard' breaking effects of $\cZ_2/SU(4)$ symmetries and it helps to reduce the fine-tuning in TH models~\cite{Katz:2016wtw}.
\een
Note that the $\mathbb{Z}_2/SU(4)$ breaking terms are important in the above potential to generate mass of the SM-like Higgs boson, which would have been a pure Goldstone boson in the absence of such terms. Moreover, these explicit symmetry breaking parameters are expected to be smaller than the symmetry preserving parameter, \ie\ $(\kappa,\sigma,\rho)\leq\lambda$, in order to not produce large radiative corrections to the SM-like Higgs mass.

We impose unitary gauge where ${\bf 6}$~Goldstone bosons are ``eaten'' by the weak gauge bosons of the SM $(W^\pm,\!Z)$ and twin  ($\hat W^\pm,\!\hat Z$) sectors. Hence, in the unitary gauge, we write the SM Higgs $H$ and twin Higgs $\widehat H$ doublets as,
\beq
H=\frac1{\sqrt2}\bpm0\\ v+h_0\epm,    \lsp \widehat H=\frac1{\sqrt2}\bpm0\\ \hat v+\hat h_0\epm.\label{H_twinH}
\eeq 
Minimization of the scalar potential~\eqref{H_AB} results in the following values of vevs, $v^2$ and $f^2\equiv v^2+\hat v^2 $,
\begin{align}
v^2=\frac{f_0^2 \big(\kappa  \lambda +\sigma  (\kappa +\lambda )\big)}{ \rho  (\kappa +\lambda )+\kappa  (\kappa +2 \lambda
   )}\,, \lsp f^2=\frac{f_0^2 \big(\kappa  (2 \lambda +\sigma )+\lambda  \rho \big)}{ \rho  (\kappa +\lambda )+ \kappa  (\kappa +2 \lambda
   )}~.
\end{align}
We define a parameter $\xi\!\equiv\! v/f$.
Roughly speaking, $\!2\xi^2$ measures the amount of fine-tuning in our model~\footnote{We do not discuss in detail, issues related to the fine-tuning in TH models but refer the interested readers to a dedicated recent paper~\cite{Contino:2017moj}, and also \cite{Craig:2015pha,Katz:2016wtw}.}. In a limit $\kappa\!\ll\! \lambda$, the $\xi^2$ can be written as,
\beq
\xi^2\simeq\frac{\sigma }{\rho }+\frac{\kappa  \big(\rho  (\lambda +\sigma )-\sigma  (2 \lambda +\sigma )\big)}{\lambda  \rho ^2}+\op\left(\frac{\kappa ^2}{\lambda^2}\right).	\label{xi_ft}
\eeq 
In order to have small fine-tuning in the model, \ie\ $\xi^2\!\ll\!1$ or $v^2\!\ll\!f^2$, one needs a tuning in the parameters $\sigma$ and $\rho$, such that $\sigma\ll \rho$ for fixed $\lambda$ and $\kappa$.  This result implies that large hard breaking effects could potentially lower the amount of fine-tuning in TH models, see also a similar discussion in \cite{Katz:2016wtw}. 

Moving on, the mass-squared matrix for the scalar fluctuations $(h_0,\hat h_0)$ is,
\beq
{\cal M}^2=\bpm 2 v^2 (\lambda +\kappa+\rho )& 2\lambda v \hat v  \\ 2 \lambda v \hat v & 2 \hat v^2(\lambda +\kappa) \epm. \label{mass_matrix}
\eeq
We diagonalize ${\cal M}^2$ by an orthogonal rotational matrix ${\cal R}$ as
\beq
{\cal M}^2_{\text{diag}}\equiv {\cal R}^{-1}{\cal M}^2{\cal R}=\bpm \mhsm^2 &0\\0&\mht^2\epm, \lsp\text{with}\hsp {\cal R}=\bpm\cos\!\alpha & ~\sin\!\alpha\\ -\!\sin\!\alpha &~\cos\!\alpha \epm,    \label{mass_matric_diag}
\eeq
where $(h,\htw)$ are the two scalar mass-eigenstates (without the subscript `0') with masses $(\mhsm^2,\mht^2)$. The two physical scalars, SM Higgs $h$ and heavy twin Higgs $\htw$, are admixture of $h_0$ and $\hat h_0$, and are defined as
\beq
h=h_0 \cos\!\alpha-\hat h_0 \sin\!\alpha,    \lsp \htw= h_0 \sin\!\alpha+\hat{h}_0\cos\!\alpha\,.\label{higgs_physical}
\eeq
The corresponding mass eigenvalues are, 
\beq
\mhsm^2=\rho \,v^2 +\left(\lambda +\kappa \right)f^2\big(1 - \Delta\big),\lsp    \mht^2=\rho\, v^2 +\left(\lambda +\kappa \right)f^2\big(1 + \Delta\big),
\eeq
where 
\beq
 \Delta^2=1-\frac{1}{\lambda+\kappa}\frac{v^2}{f^2}\left[\frac{4(\kappa^2+2\lambda\kappa)}{\lambda+\kappa}\left(1+\frac{v^2}{f^2}\right)+2\rho\left(1+2\frac{v^2}{f^2}\right)+\rho^2\frac{v^2}{f^2}\right].
\eeq
The mixing angle is given by
\beq
\tan(2\alpha)=\frac{2 \xi  \sqrt{1-\xi ^2}}{ \left(1 -(2 +\rho/\lambda ) \xi^2\right)+\left(1-2 \xi^2\right)\kappa/\lambda }~.
\eeq
For more intuition of the above results, we expand in the limit $\xi^2\!=\!v^2/f^2\ll 1$. At the leading order in this approximation, we get the mass eigenvalues as, 
\begin{align}
\mhsm^2&=2 \rho v^2 +\frac{2v^2 (\kappa^2 +2 \kappa\lambda )}{\kappa +\lambda },        
&\mht^2&=2 f^2 (\kappa +\lambda )-\frac{2 v^2 (\kappa^2+2\kappa \lambda )}{\kappa +\lambda }\,. \label{mth_ov2f2}
\end{align}
The above expressions manifestly show that in the limit $\xi^2\!\ll\! 1$ and fixed $v^2$, the SM Higgs mass is directly related to hard breaking parameter $\rho$. Therefore, the hard breaking parameter $\rho$ would be severely constrained by the SM Higgs mass. However, it is possible to have sizable $\rho$, in order to reduce the fine-tuning as mentioned above in Eq.~\eqref{xi_ft}, but at the price of reducing $\kappa$ to get the SM Higgs mass correct, see also~\cite{Katz:2016wtw}.
Furthermore, in the same approximation $\xi^2\!\ll\! 1$, the mixing angle can be written as, 
\beq
\sin^2\!\alpha=\frac{\lambda^2 }{(\lambda+\kappa )^2}\frac{v^2}{f^2}+\op\Big(\frac{v^4}{f^4}\Big),
\eeq
which shows a dependence of $\sin^2\!\alpha\simeq v^2/f^2$ similar to composite Higgs model.  

Before going ahead, we would like to comment on the most relevant scalar sector couplings, the triple Higgs couplings $g_{hhh}$ and $g_{\hat hhh}$, which are given as 
\begin{align}
g_{hhh}&=6\Big[v\cos\!\alpha\big(\lambda+(\kappa+\rho)\cos^2\!\alpha\big)-\hat v\sin\!\alpha\big(\lambda+\kappa\sin^2\!\alpha\big)\Big],    \label{ghhh}\\
g_{\hat h hh}&=2\Big[v\sin\!\alpha\big(\lambda+3(\kappa+\rho)\cos^2\!\alpha\big)+\hat v\cos\!\alpha\big(\lambda+3\kappa\sin^2\!\alpha\big)\Big].    \label{gthhh}
\end{align}
Trilinear Higgs coupling of the SM-like Higgs $g_{hhh}$ is a very crucial indirect probe of new physics at the current and future colliders. Any deviation from its corresponding SM value, 
\beq
g^{\textsc{sm}}_{hhh}=\frac{3 m_h^2}{v},
\eeq
would be a signal of new physics. At the high-luminosity LHC (HL-LHC), \ie\ $\sqrt s\!=\!14\tev$ with 3000 fb$^{-1}$, it is expected to achieve sensitivity of $|1\!-g_{hhh}/g^{\textsc{sm}}_{hhh}|\!\sim\!0.5$,~\cite{Dawson:2013bba}. However, as we see in the following sections, in most of TH parameter space trilinear Higgs coupling is very mildly constraining. On the other hand, the triple Higgs coupling $g_{\hat h hh}$ is crucial for the decay of heavy twin Higgs to a pair of SM Higgs. 

\subsection{Physical Basis for Twin Higgs models}
Twin Higgs effective scalar potential~\eqref{Veff_AB} contains five real parameters, $f_0,\,\lambda,\, \kappa,\,\sigma$ and $\rho$. For phenomenological analysis, it is more convenient to choose a basis where these parameters can be expressed in more physical observables. Physical observables for linear TH models are the SM Higgs mass~$m_h$, heavy twin Higgs mass~$\mht$, SM vev~$v$, and twin vev~$\hat v$ (or equivalently $f$). We fix the SM Higgs mass $m_h\!=\!125\gev$ and its vev $v\!=\!246\gev$. This leaves us with three free parameters $\mht$, $f$, and hard breaking parameter $\rho$. Since we require $\rho\leq\lambda$, therefore, it is instructive to define $\tilde\rho$ as  
\beq
\tilde\rho\equiv \rho/\lambda,    \lsp\text{such that}\lsp |\tilde\rho|\lsim1\,.
\eeq
In the following sections we work in $(\mht\!-\!f\!/\!v)$ plane and constraint the parameter space through direct searches of the heavy twin Higgs and indirect limits from the SM Higgs data at the LHC. The trade of potential parameters in the `physical TH basis' is as, 
\beq
\underbrace{f_0,\,\lambda,\, \kappa,\,\sigma,\,\rho}_{\rm TH~gauge~basis} \hsp\Longleftrightarrow \hsp \underbrace{v,\,f,\,m_h,\,\mht,\,\tilde\rho}_{\rm TH~physical~basis}~.
\eeq
Advantage of the above parametrization is that it leaves the hard breaking parameter $\tilde \rho$ as a free parameter, which gives us a handle on hard breaking effects in the model.
Potential parameters have the following values in the physical basis:
\begin{align}
f_0&=f \sqrt{1+\frac{\kappa}{\lambda} \big(1-\xi ^2\big)}~,        \Lsp \sigma=\frac{\lambda  \left(\lambda  \tilde\rho \xi ^2-\kappa  \left(1-2 \xi
   ^2\right)\right)}{\lambda +\kappa (1- \xi ^2)}~,        \label{f0-sigma_phy}\\
\lambda&=\frac{\big(\mht^2+m_h^2\big)\tilde\rho(1-2\xi^2)}{4f^2\left(1+\tilde\rho\xi^2(1-\xi^2)\right)}+\frac{\widehat\Delta}{4f^2\xi\sqrt{1-\xi^2}\left(1+\tilde\rho\xi^2(1-\xi^2)\right)}~,        \label{lamb_phy}\\
\kappa&=\frac{\big(\mht^2+m_h^2\big)\big(2+\tilde\rho^2\xi^2-\tilde\rho(1-2\xi^2)\big)}{4f^2\left(1+\tilde\rho\xi^2(1-\xi^2)\right)}    -\frac{(1+\tilde\rho\xi^2)\,\widehat\Delta}{4f^2\xi\sqrt{1-\xi^2}\left(1+\tilde\rho\xi^2(1-\xi^2)\right)}\, ,        \label{kapa_phy}
\end{align}
where
\beq
\widehat \Delta\equiv \sqrt{\big[4(\mht^2+m_h^2)^2    +(\mht^2-m_h^2)^2\tilde\rho^2\big]\xi^2(1-\xi^2)-4\mht^2m_h^2}~.
\eeq
In our phenomenological analysis in Secs.~\ref{Heavy Higgs Phenomenology in the Mirror Twin Higgs}  and \ref{Heavy Higgs Phenomenology in the Fraternal Twin Higgs}, we require the following conditions on potential parameters in order to satisfy unitarity and perturbativity constraints, 
\beq
\lsp 0<\lambda\leq 4\pi, \lsp |\sigma|\leq \lambda, \lsp |\kappa|\leq \lambda,\lsp |\rho|\leq \lambda, \hsp{\text{and}}\hsp f_0^2>0,    \label{physical}
\eeq
along with the SM Higgs mass $m_h\!=\!125\gev$ and its vev $v\!=\!246\gev$.
If any (or more than one) of the above conditions is not met then the parameter space would be called ``unphysical''.

\subsection{Electroweak Precision Constraints in Twin Higgs Models}
\label{ewpt}
In this subsection, we briefly comment on the constraints in the TH models from the electroweak precision data when the TH symmetry breaking is linearly realized and a relatively light radial mode (twin Higgs) is present in the effective theory. In particular, the electroweak oblique parameters $S$ and $T$ receive infrared contributions --- due to the mixing of the SM-like Higgs with the heavy states and the reduced coupling of the SM-like Higgs to the SM gauge bosons --- which are proportional to $(1-a^2)\ln(\Lambda^2/m_h^2)$, where $a\!\equiv\!g_{hVV}/g^{\rm SM}_{hVV}$ and $\Lambda$ is the scale associated with the heavy states~\cite{Espinosa:2012im}. In the linear TH models, $a=g_h=\cos\alpha$ (see Fig.~\ref{fig:couplings}) and the cutoff scale $\Lambda$ is the heavy twin Higgs mass, i.e. $\Lambda\!=\!\mht$. Hence, the shifts of the oblique parameters $S$ and $T$ due to the SM-like Higgs and heavy twin Higgs in the TH models are~\cite{Craig:2015pha,Espinosa:2012im}, 
\begin{align}
\Delta S\approx \frac{1}{12}\frac{\sin^2\!\alpha}{\pi}\,\ln\bigg(\frac{\mht^2}{m_h^2}\bigg), \Lsp \Delta T\approx -\frac{3}{16}\frac{\sin^2\!\alpha}{\pi\cos^2\!\theta_W}\,\ln\bigg(\frac{\mht^2}{m_h^2}\bigg).
\end{align}
\begin{figure}[t]
\centering
\includegraphics[width=0.5\textwidth]{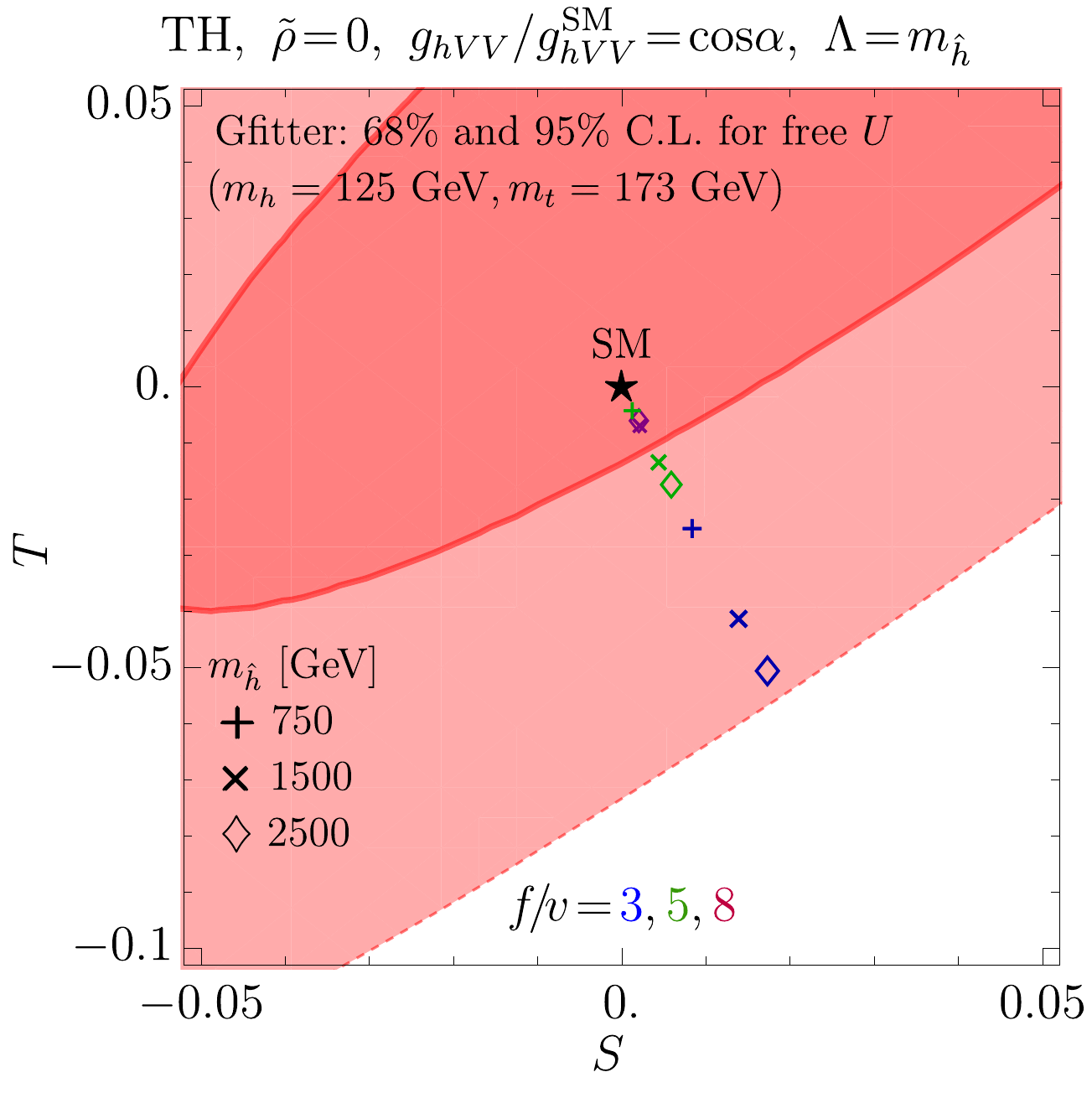}\includegraphics[width=0.5\textwidth]{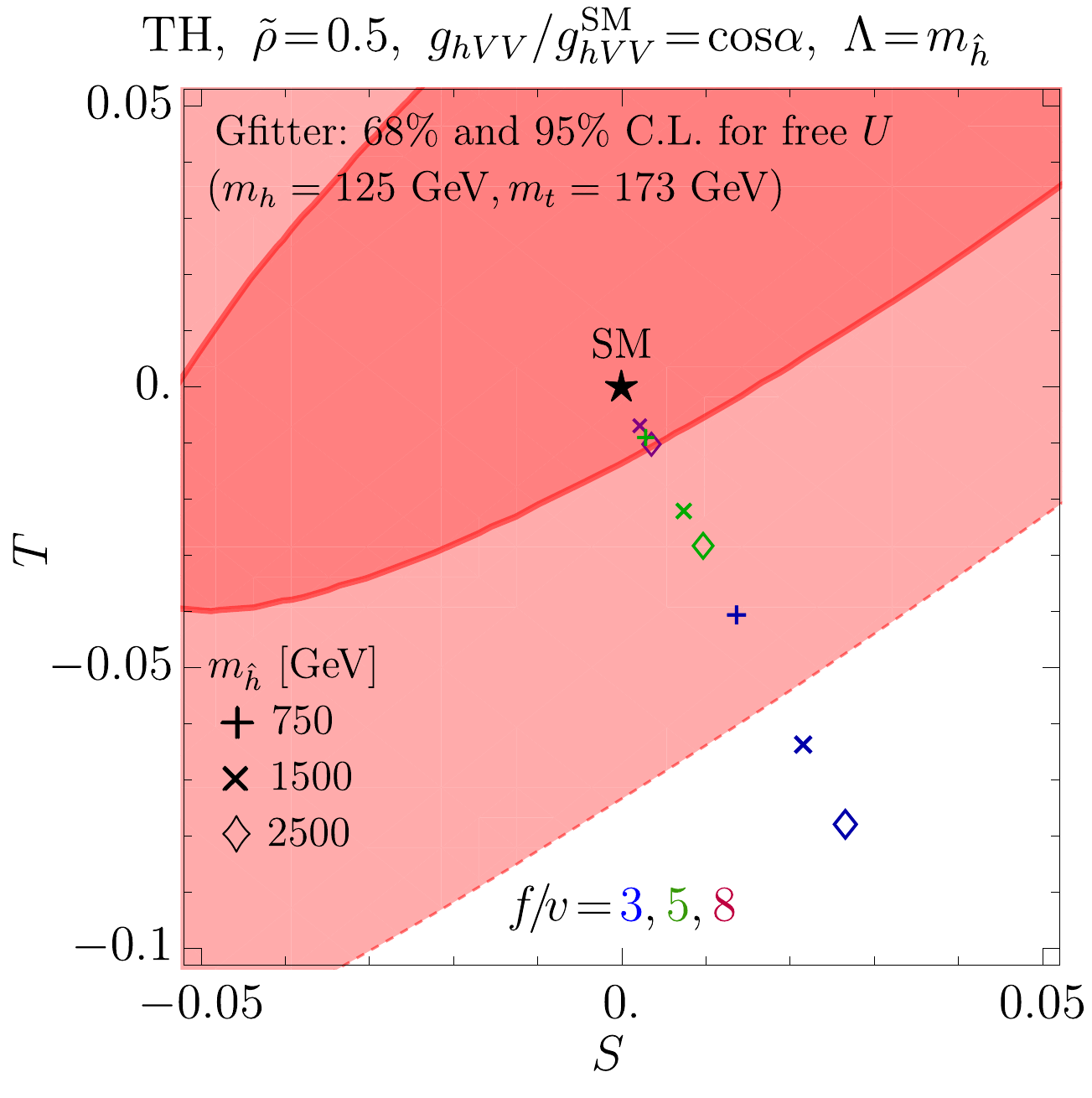}
\caption{Magnified $68\% (\!\rm\it red\!)$ and $95\% (\!\text{\it light-red})$ C.L. contours of the oblique parameters $S,T$ showing contributions of the scalar sector of the TH models for $\tilde \rho\!=\!0$ (left) and $\tilde \rho\!=\!0$ (right). We consider the twin Higgs mass $\mht\!=\!(750,1500,2500)\!\gev$ denoted as $(+,\times,\diamondsuit)$, respectively, for the color-coded values of $f\!/\!v\!=\!3 (\!\rm\it blue\!),5 (\!\rm\it green\!),8 (\!\rm\it purple\!)$. }
\label{fig:stu}
\end{figure}
In Fig.~\ref{fig:stu}, we employ Gfitter values for $S\!=\!0.05\!\pm\!0.11$ and $T\!=\!0.09\!\pm\!0.13$ with free $U$ for the SM Higgs $m_h\!=\!125\gev$~\cite{Baak:2014ora} to construct $68\% (\!\rm\it red\!)$ and $95\% (\!\text{\it light-red})$ C.L. $S\!-\!T$ contours. Twin Higgs contributions to the oblique parameters are shown for $\mht\!=\!(750,1500,2500)\!\gev$ as $(+,\times,\diamondsuit)$, respectively, for the color-coded values of $f\!/\!v=$~3~({\it blue}), 5~({\it green}), 8~({\it purple}). The contribution of a pure SM Higgs of mass $m_h\!=\!125\gev$ is shown as a black star $\star$. The left (right) plot considers hard breaking contributions $\tilde\rho\!=\!0(0.5)$ in the TH effective potential. Note that the presence of hard breaking contributions $\tilde\rho\!>\!0~(\tilde\rho\!<\!0)$ slightly worsens (improves) the electroweak precision constraints as compared to the $\tilde\rho\!=\!0$ case. This is mainly due to the fact that the hard breaking effects are proportional to the mixing angle $\alpha$, see also below. 
There are two interesting features we observe from Fig.~\ref{fig:stu}: ({\it i}) for a fixed value of $f/v$, the TH contributions to the oblique parameters $S,T$ are smaller for lighter twin Higgs, and ({\it ii}) for a fixed twin Higgs mass, the oblique parameters $S,T$ receive contribution proportional to $v^2\!/\!f^2$, this feature is similar to composite Higgs models. In the following sections for the phenomenological analysis, we implement the electroweak precision constraints at $95\%$ C.L on the TH parameter space.

\section{Heavy Higgs Phenomenology in Mirror Twin Higgs} 
\label{Heavy Higgs Phenomenology in the Mirror Twin Higgs} 
The most general effective renormalizable Lagrangian for the MTH model is~\footnote{We have not included a renormalizable $\cZ_2$ soft breaking term involving mixing of the SM hypercharge $U(1)_Y$ and twin hypercharge $U(1)_{\hat Y}$, \ie\ $\varepsilon B_{\mu\nu}\hat B^{\mu\nu}$. It provides a vector portal between the SM and twin sectors, see ~\cite{Saereh:2017tp}.}
\beq
\cL_\mth=\cL_{321}+\hat\cL_{\hat3\hat2\hat 1}-V_{\eff}(H,\widehat H),    \label{lang_mth}
\eeq  
where $\cL_{321}$ represents full SM Lagrangian with $SU(3)_c\!\times\! SU(2)_L\!\times\! U(1)_Y$ gauge group excluding the SM Higgs potential and $\hat\cL_{\hat3\hat2\hat 1}$ is the corresponding twin sector Lagrangian. The effective scalar potential $V_\eff(H,\widehat H)$ for the MTH model is given in Eq.~\ref{Veff_AB}. Notable features of the MTH model are:
\bit\itemsep0em
\item[$\diamond$] Gauge structure of twin Lagrangian $\hat\cL_{\hat3\hat2\hat 1}$ is a mirror copy of the SM Lagrangian~$\cL_{321}$, and the twin sector particle content is the same as that of the SM. 
\item[$\diamond$] Twin sector gauge and Yukawa couplings are almost equal to that of the SM, \ie\
\beq
\hat g_s\simeq g_s, \lsp \hat g\simeq g,\lsp \hat g^\p\simeq g^\p,\lsp \hat y_{\hat f}\simeq y_{f}\,,
\eeq
where $g_s,g$ and $g^\p$ denote the $SU(3)_c,SU(2)_L$ and $U(1)_Y$ gauge couplings, respectively, whereas $y_f$ are the fermion Yukawa couplings. 
This is required to cancel the quantum corrections to the SM Higgs mass. 
\item[$\diamond$] Masses of the twin sector gauge bosons $m_{\hat V}$ and fermions $ m_{\hat f}$ scale universally w.r.t. the SM counterparts as, 
\beq
m_{\hat V} =\frac{\hat v}{v}m_{V}\,,~\lsp m_{\hat f} =\frac{\hat v}{v}m_{f},
\eeq
where $V\!=\!(W^\pm,Z)$ and $f$ denotes the SM fermions (twin sector are with a {\it hat}). 
\item[$\diamond$] We assume a mechanism for twin neutrino mass generation exits, see \eg\ \cite{Chacko:2016hvu,Csaki:2017spo}. However, neutrinos play almost no role in the phenomenology of TH scalar sector so they are irrelevant to our discussion. 
\eit
We collect the most relevant Feynman rules for the scalar sector of the MTH model in Fig.~\ref{fig:couplings} (Appendix~\ref{Partial decay widths of heavy twin Higgs}), where apart from tree-level vertices for the SM and twin-sector massive gauge bosons and fermions, we also include loop induced vertices for massless gauge bosons of the two sectors. Note that the SM-like Higgs couplings to the SM sector particles are suppressed by $g_h\!=\!\cos\alpha$ to their respective SM values, which is a common feature of BSM scenarios where the SM Higgs mixes with another scalar with a mixing angle $\alpha$. The SM Higgs coupling measurements at the LHC put severe constraints on the mixing angle and exclude some of the parameter space of the MTH model.

We collect formulae for partial decay widths of heavy twin Higgs to the SM and twin states in Appendix~\ref{Partial decay widths of heavy twin Higgs}.
For computing heavy twin Higgs cross-sections, we adopt a strategy similar to the one outlined in~\cite{Ahmed:2015uqt}. For different initial (production) and final (decay) states, the cross-section of heavy twin Higgs is parametrized as: 
\beq
\sigma^{YY}_{\hat h\to XX}\!\equiv\!\sigma(YY\to\hat h)\!\cdot\!{\cal B}(\hat h\to XX)=\sigma^{\text{SM}}(YY\to h)\big\vert_{m_h\!=\!m_{\hat h}}~{\cal C}_{\hat h YY}^2  ~{\cal B}(\hat h\to XX),
\eeq
where $\sigma^{\text{SM}}(YY\!\to\! h)$ is the SM-like Higgs production cross-section calculated at the heavy twin Higgs mass for $YY\!=\!(gg,VV)$.  The SM Higgs cross-sections $\sigma^{\text{SM}}(h\to YY )|_{m_h\!=\!m_{\hat h}}$ are obtained from Higgs Cross-section Working Group~\cite{deFlorian:2016spz}. Above ${\cal B}(\hat h\to XX)$ is the twin Higgs branching fraction to $XX$ final state, whereas, effective coupling ${\cal C}_{\hat h YY}$ is defined as, 
\beq
{\cal C}^2_{\hat h YY} \equiv \frac{\sigma(YY \to\hat h)}{\sigma^{\text{SM}}(YY \to h)|_{m_h\!=\!m_{\hat h}}}=\frac{\Gamma(\hat h\to YY )}{\Gamma^{\text{SM}}(h\to YY )|_{m_h\!=\!m_{\hat h}}}~,
\eeq
where $\Gamma^{\text{SM}}(h\to YY )|_{m_h\!=\!m_{\hat h}}$ is the SM Higgs partial decay width evaluated at the heavy Higgs mass and $\Gamma(\hat h\to YY )$ is the twin Higgs decay width, see Appendix~\ref{Partial decay widths of heavy twin Higgs}. For $YY=gg$ and $VV$, one finds the effective twin Higgs coupling,
\beq
{\cal C}^2_{\hat h gg} ={\cal C}^2_{\hat h VV} = \big|g_{\hat h}\big|^2=\sin^2\!\alpha~,
\eeq
which is a universal rescaling of heavy twin Higgs production at the LHC and future colliders. 
Note that the main production channel for the heavy twin Higgs is gluon-gluon fusion ($gg\textsc f$), whereas, the vector-boson fusion (\textsc{vbf}) is subdominant in most of the parameter space of TH models.
Similarly, for the SM Higgs with $m_h\!=\!125\gev$, the cross-section times branching fractions are given as, 
\beq
\sigma^{YY}_{h\to XX}\!\equiv\!\sigma(YY\to h)\!\cdot\!{\cal B}(h\to XX)=\sigma^{\text{SM}}(YY\to h)~{\cal C}_{h YY}^2  ~{\cal B}(h\to XX),
\eeq
where for $YY=(gg,VV)$ the effective SM Higgs coupling is
\[
{\cal C}^2_{h gg} ={\cal C}^2_{h VV} = \big|g_{h}\big|^2=\cos^2\!\alpha~.
\]
It is convenient to define SM Higgs signal-strength $\mu^{YY}_{h\!\to\!XX}$ as, 
\beq
\mu^{YY}_{h\!\to\!XX}\equiv \frac{\sigma(YY\to h)\!\cdot\!{\cal B}( h\to XX)}{\sigma^{\rm SM}(YY\to h)\!\cdot\!{\cal B}^{\rm SM}(h\to XX)}={\cal C}_{h YY}^2\,\frac{{\cal B}(h\to XX)}{{\cal B}^{\rm SM}(h\to XX)}\,.
\eeq
The SM Higgs signal-strength measurements in many production and decay channels have been measured with unforeseeable precision at the LHC run-1~\cite{Khachatryan:2016vau} and they provide stringent constraints on BSM models. 

For numerical analysis of the twin Higgs phenomenology, we employ a modified version of {\tt HDECAY} code~\cite{Djouadi:1997yw} incorporating the twin sector, where all the appropriate QCD and electroweak corrections are included in the SM as well as in the twin sector. Note that in the MTH running of twin gauge couplings is similar to those of the SM. We also assume that at the cutoff scale $\Lambda$, the SM and twin gauge coupling as well the Yukawa couplings are exactly equal. Moreover, the heavy twin Higgs (and the SM-like Higgs) decays to the twin states are invisible at the colliders.  

\begin{figure}[t]
\centering
\includegraphics[width=0.49\textwidth]{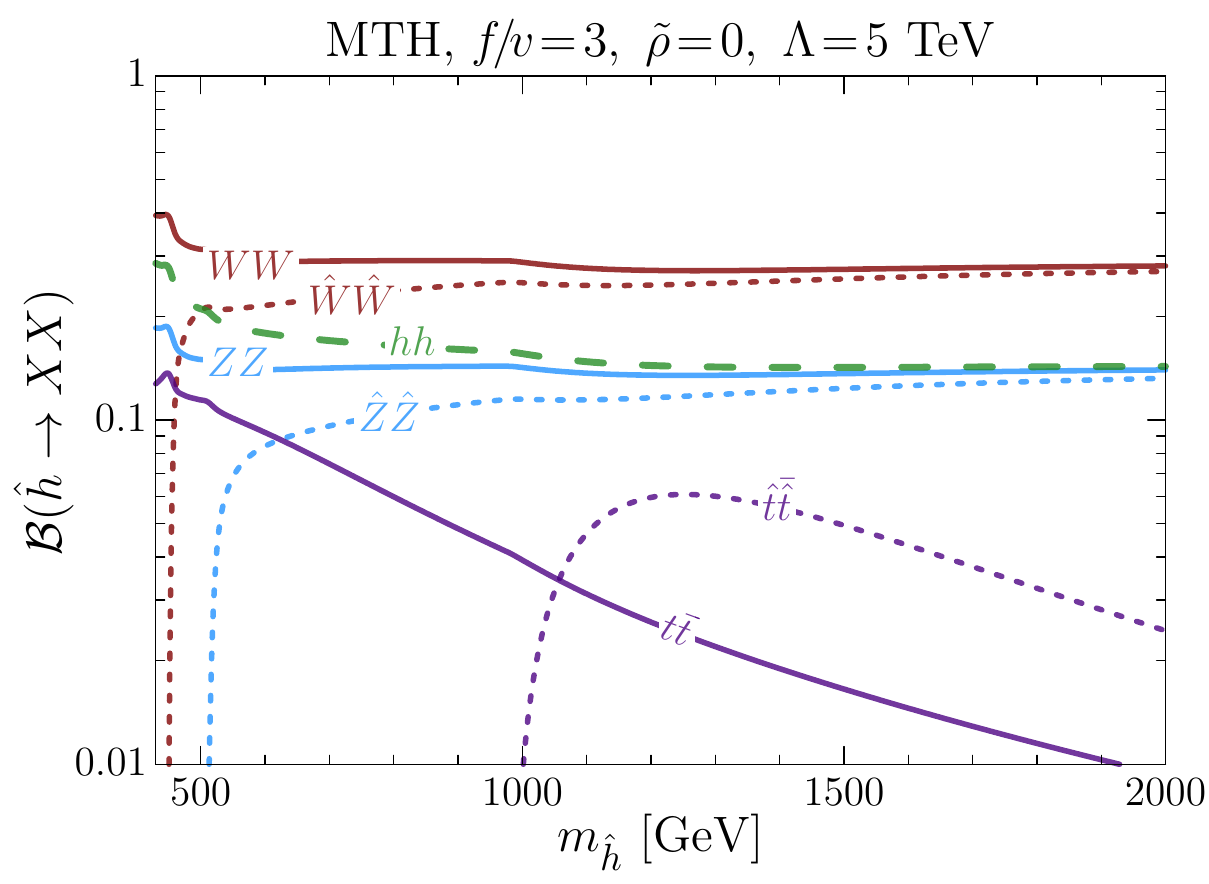}\!
\includegraphics[width=0.49\textwidth]{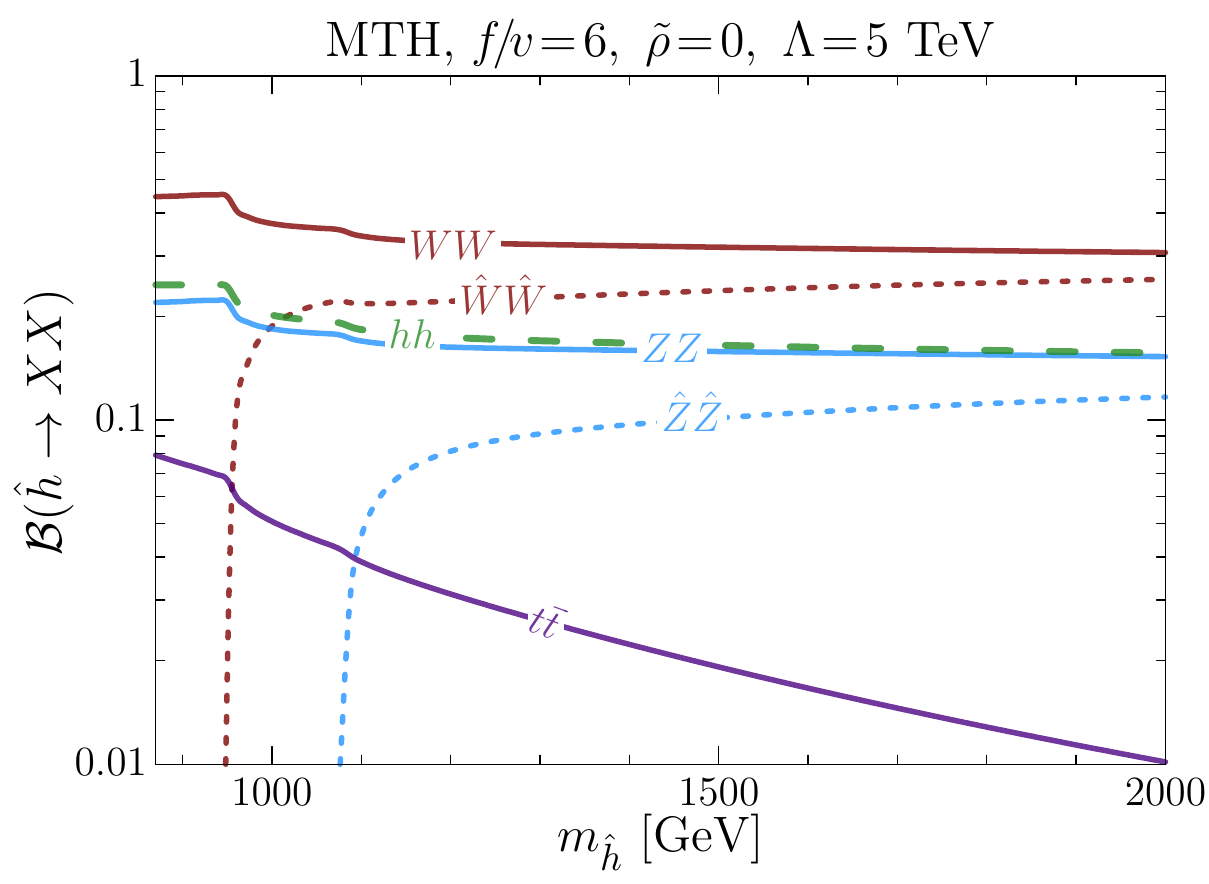}
\caption{These plots show heavy twin Higgs branching ratios in the MTH to different final states as a function of its mass. We take $f\!/\!v\!=\!3$ (left-panel) and $6$ (right-panel) for $\tilde\rho\!=\!0$ and $\Lambda\!=\!5\tev$.}
\label{fig:mthbrsr0}
\end{figure}

In Fig.~\ref{fig:mthbrsr0} we show branching fractions (larger than a percent) of heavy twin Higgs for different final states of the SM and twin (invisible) sector as a function of the heavy Higgs mass for $f\!/\!v\!=\!3$ (left-panel) and 6 (right-panel), and $\tilde\rho\!=\!0$. Notice the lower mass of the heavy Higgs is shifted to higher values as $f\!/\!v$ increases. This is due to the fact that with the increase in $f\!/\!v$ the lower twin Higgs masses do not reconcile with the SM Higgs mass $125\gev$. For a given production mode (\eg\ gluon fusion), the relative magnitudes of these branching ratios determine the relative rates for the various final states. In the absence of the $\cZ_2/SU(4)$ hard breaking term, \ie\ $\tilde \rho=0$, it is clear from Fig.~\ref{fig:mthbrsr0} the heavy Higgs branching fractions to the SM vector bosons $W^\pm, Z$, Higgs $h$, and the twin vector bosons $\hat W^\pm, \hat Z$, are the most dominant channels. Moreover, for $\tilde \rho=0$ we find,
\beq
{\cal B}(\hat h\to hh)\simeq{\cal B}(\hat h\to ZZ)\simeq\tfrac12{\cal B}(\hat h\to WW)\simeq{\cal B}(\hat h\to \hat Z\hat Z)\simeq\tfrac12{\cal B}(\hat h\to \hat W\hat W),
\eeq
which is a clear manifestation of the Goldstone boson equivalence theorem. Since the longitudinal modes of massive vector bosons and the SM Higgs are Goldstone modes of the TH spontaneous symmetry breaking. Since the radial mode couples universally to all the Goldstone modes, therefore, one would expect that the branching ratio of heavy twin Higgs in the visible ($W^\pm,Z,h$) sector would be $\sim\!4/7$ and that in the invisible ($\hat W^\pm,\hat Z$) sector would be $\sim\!3/7$, see also \cite{Buttazzo:2015bka}. This gives a profound prediction for the TH models that the decay rates of heavy Higgs to the SM Higgs and vector bosons (say $ZZ$) are equal. Such an observation at the LHC and/or future colliders would provide a clear distinction between TH and other BSM models. 
\begin{figure}[t]
\centering
\includegraphics[width=0.49\textwidth]{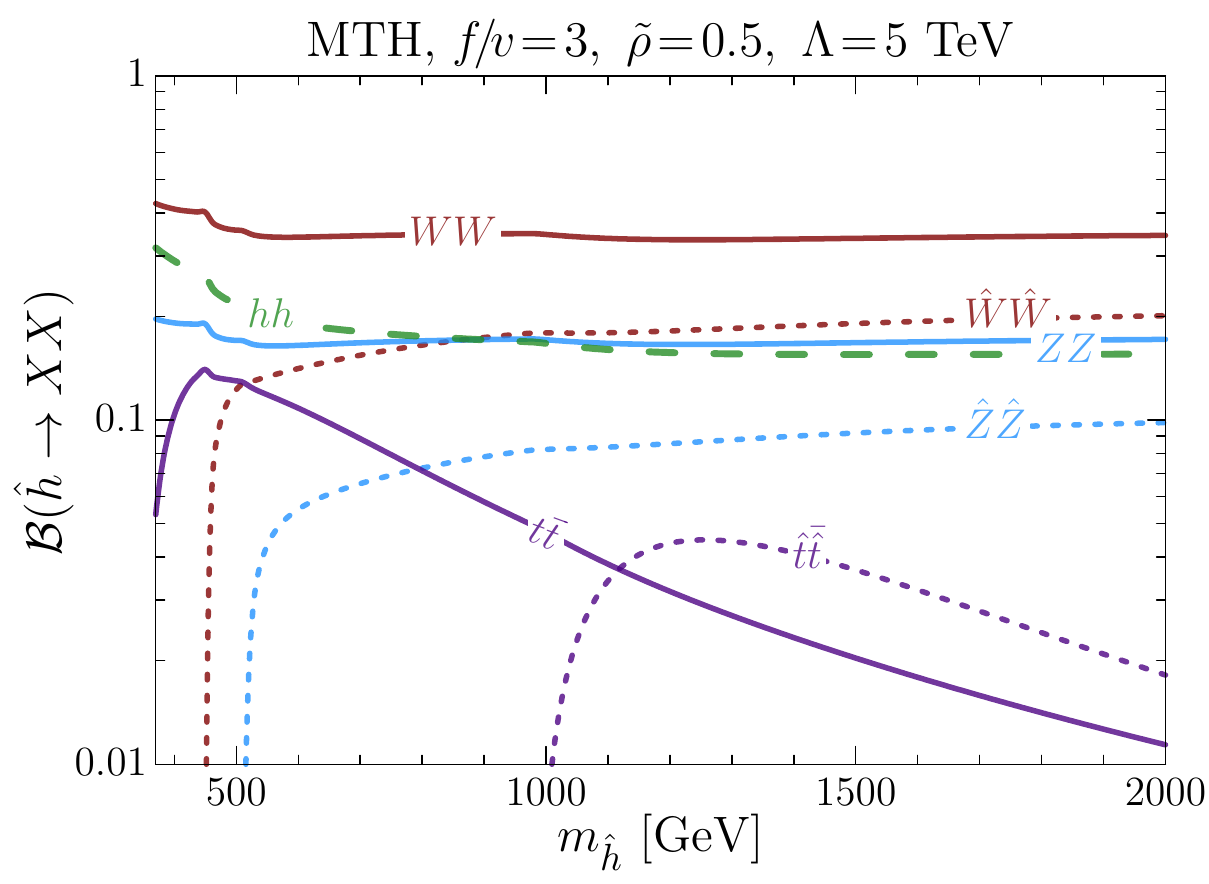}\!
\includegraphics[width=0.49\textwidth]{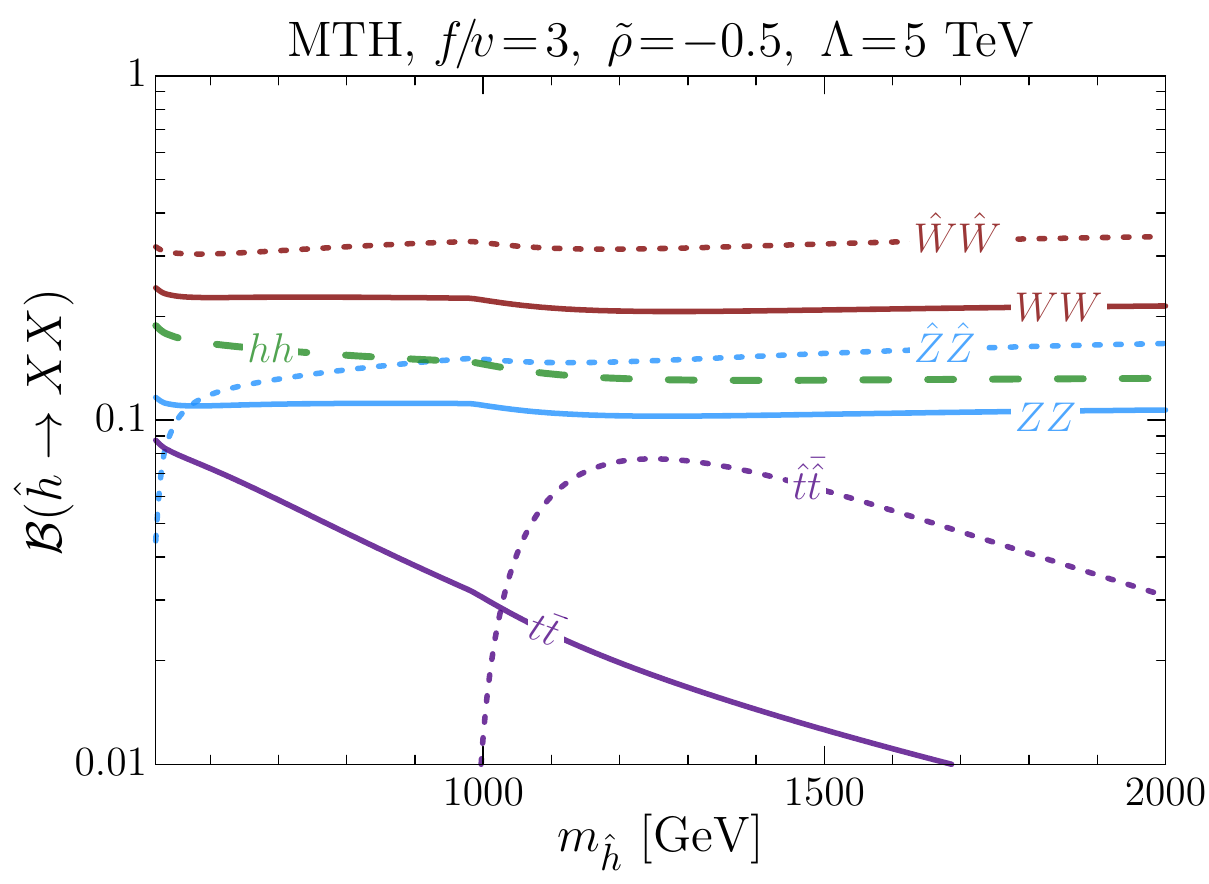}
\caption{Same as Fig.~\ref{fig:mthbrsr0} but for explicit hard breaking parameter $\tilde\rho=0.5$ (left) and $-0.5$ (right).}
\label{fig:mthbrsr05}
\end{figure}

On the other hand, in the presence of explicit $\mathbb{Z}_2/SU(4)$ hard breaking $\rho$-term in the TH scalar potential \eqref{Veff_AB}, the above-mentioned universality among the interactions of heavy twin Higgs to the SM Higgs/vector bosons and twin vector bosons would break. To see this explicit symmetry breaking effects, in Fig~\ref{fig:mthbrsr05} we plot the twin Higgs branching fractions for $\tilde \rho\!=\!0.5$ and $-0.5$, where $\tilde \rho\equiv \rho/\lambda$. As we are working in EFT framework where the sign and/or the size of explicit hard braking term is a model-dependent quantity so we consider both sign possibilities, see for instance~\cite{Katz:2016wtw}. Notice for the case of positive $\tilde \rho$ (Fig~\ref{fig:mthbrsr05} left-panel), there is a sharp increase of SM branching fractions, whereas this behavior is reversed for the case of negative $\tilde\rho$ (right-panel), as compared to Fig~\ref{fig:mthbrsr0} (left-panel) where $\tilde\rho\!=\!0$. The reason for this is the effects of explicit breaking on mixing angle $\alpha$, which increases for positive $\tilde\rho$ and vice versa (see also~\eqref{mth_ov2f2}). Hence, $\tilde\rho\!>\!0$ enhances the rates of heavy twin Higgs to the SM sector, whereas $\tilde\rho\!<\!0$ suppresses those rates. As a consequence, the LHC (and/or future colliders) would be able to probe more region of TH parameter space with $\tilde \rho\!>\!0$ than for $\tilde \rho\!<\!0$ (see below).

\begin{figure}[t]
\centering
\includegraphics[width=0.49\textwidth]{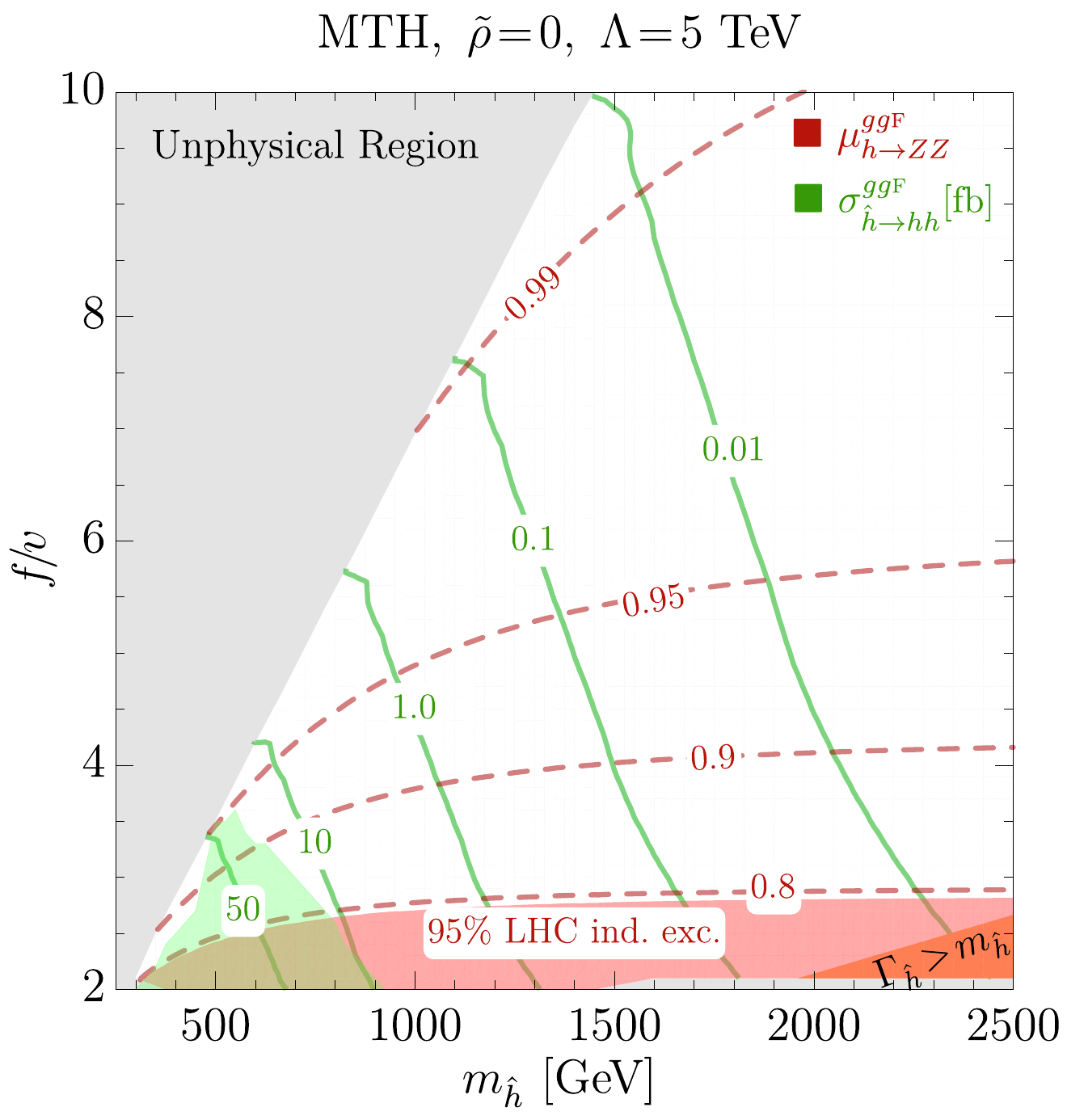}
\includegraphics[width=0.49\textwidth]{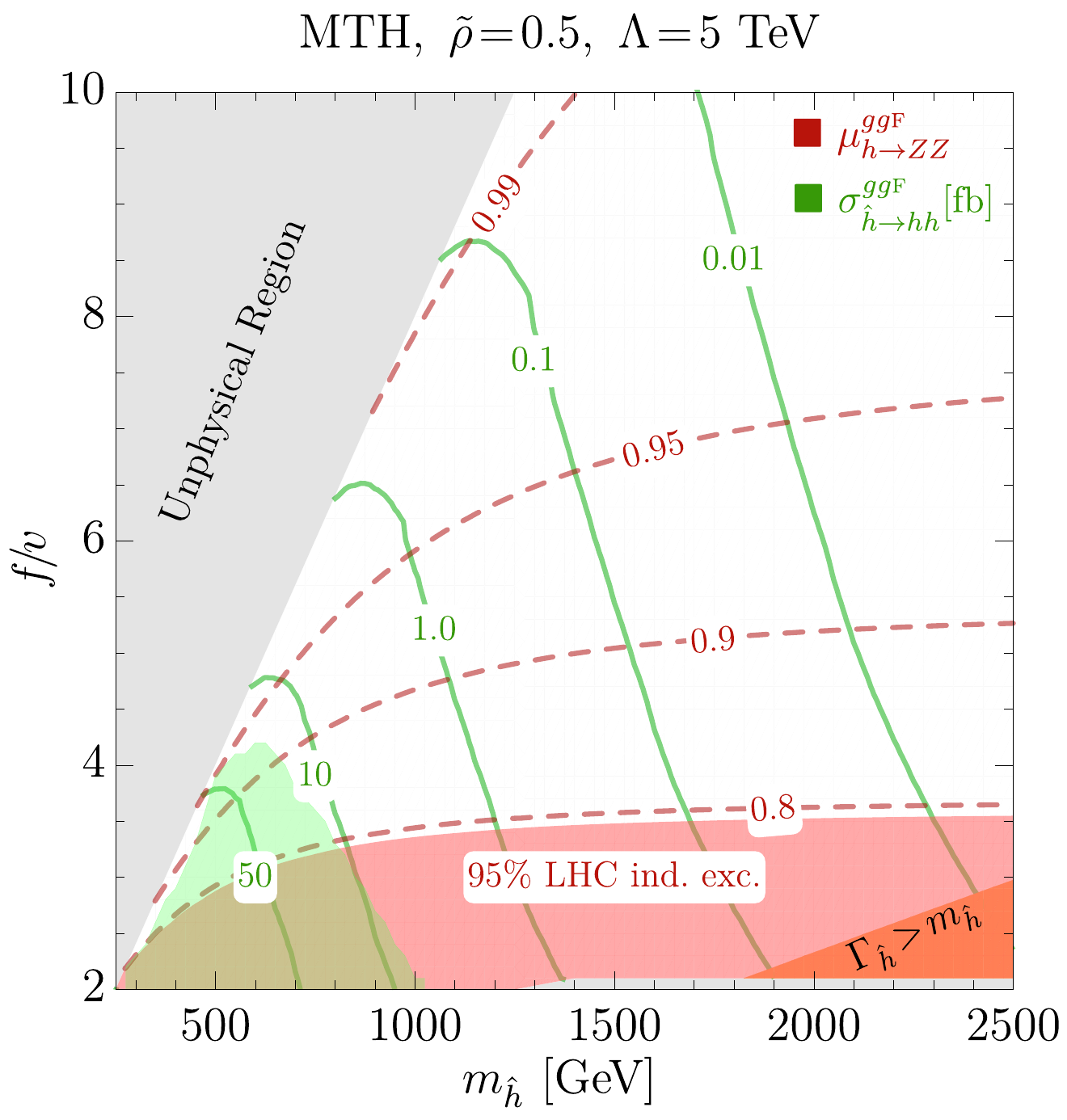}
\caption{These plots show parameter space in $\mht\!-\!f\!/\!v$ plane for the MTH model. The contours of heavy twin Higgs cross-section to a pair SM Higgs bosons $\sigma^{gg\textsc{f}}_{\hat h\!\to\! hh}$~[fb] (solid green) and SM Higgs signal-strength $\mu^{gg\textsc{f}}_{h\!\to\!ZZ}$ (dashed red) are shown for $\tilde \rho\!=\!0$ (left) and $\tilde \rho\!=\!0.5$ (right). Shaded regions read as: unphysical parameters~(gray), $\Gamma_{\hat h}\!>\!\mht$~(orange), the projected reach for direct searches $\hat h\!\to\! hh$ at the HL-LHC~(light-green), and exclusion by the SM Higgs signal-strength measurements at the LHC run-1~(red).}
\label{fig:mthfhhr0}
\end{figure}

In Fig.~\ref{fig:mthfvvr0} we show parameter space of the MTH model in $\mht\!-\!f\!/\!v$ plane with contours of heavy twin Higgs cross-sections to a pair of SM vector bosons \ie\ $\sigma^{gg\textsc{f}}_{\hat h\!\to\! ZZ}\!\equiv\!\sigma(gg\!\to\! \hat h)\!\cdot\!{\cal B}(\hat h\!\to\! ZZ)$ (solid blue curves), and SM Higgs signal-strength $\mu^{gg\textsc{f}}_{h\!\to\!ZZ}$ (dashed red curves). The cross-section is calculated at the LHC with the center-of-mass energy $14\tev$. The left and right plots in Fig.~\ref{fig:mthfvvr0} are for hard breaking  parameter $\tilde \rho\!=\!0$ and $\tilde \rho\!=\!0.5$, respectively.  Similarly, Fig.~\ref{fig:mthfhhr0} presents contours of heavy twin Higgs cross-section to a pair of SM Higgs bosons, $\sigma^{gg\textsc{f}}_{\hat h\!\to\! hh}\!\equiv\!\sigma(gg\!\to\! \hat h)\!\cdot\!{\cal B}(\hat h\!\to\! hh)$ (solid green curves).
\begin{figure}[t]
\centering
\includegraphics[width=0.49\textwidth]{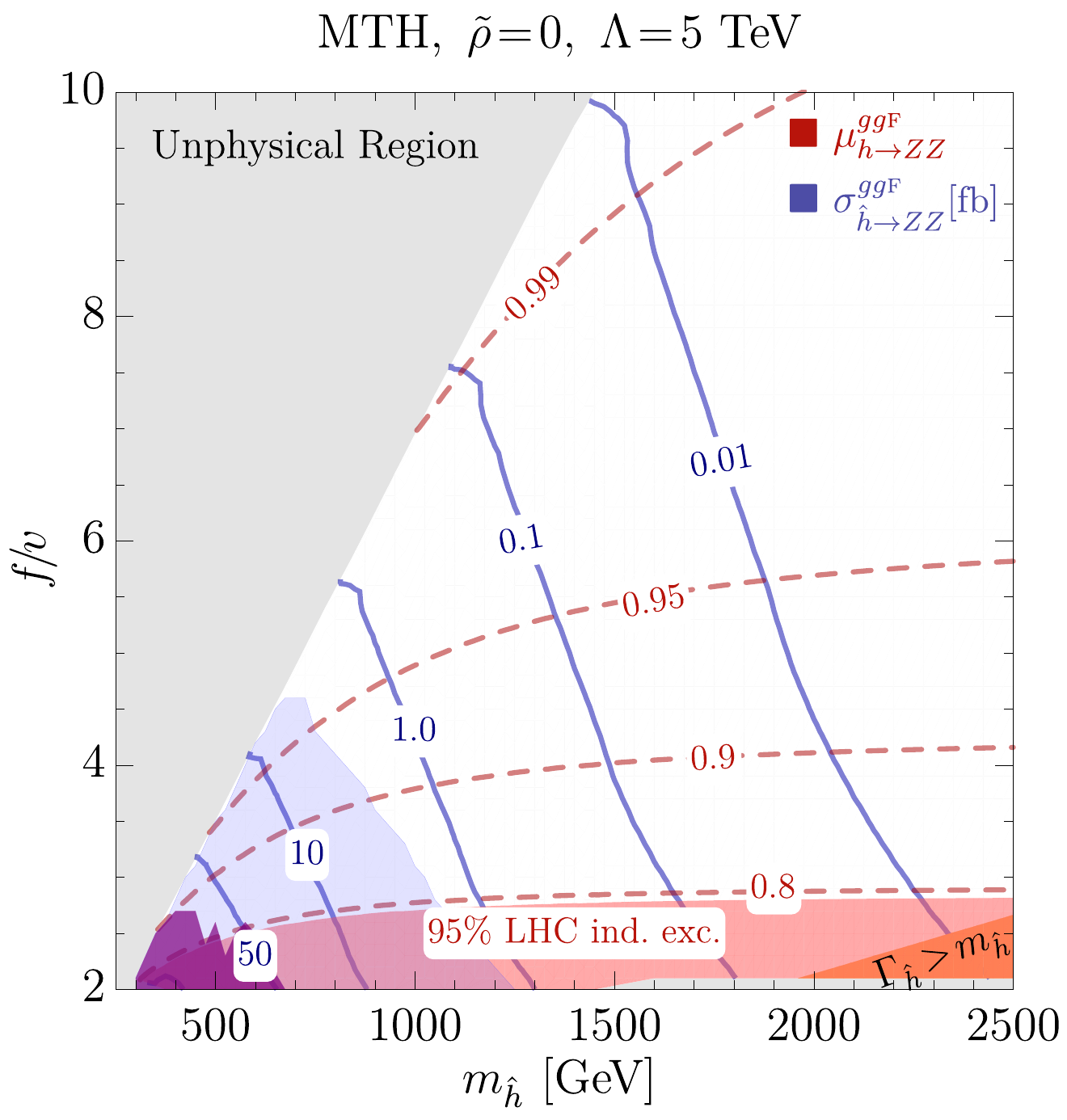}\!
\includegraphics[width=0.49\textwidth]{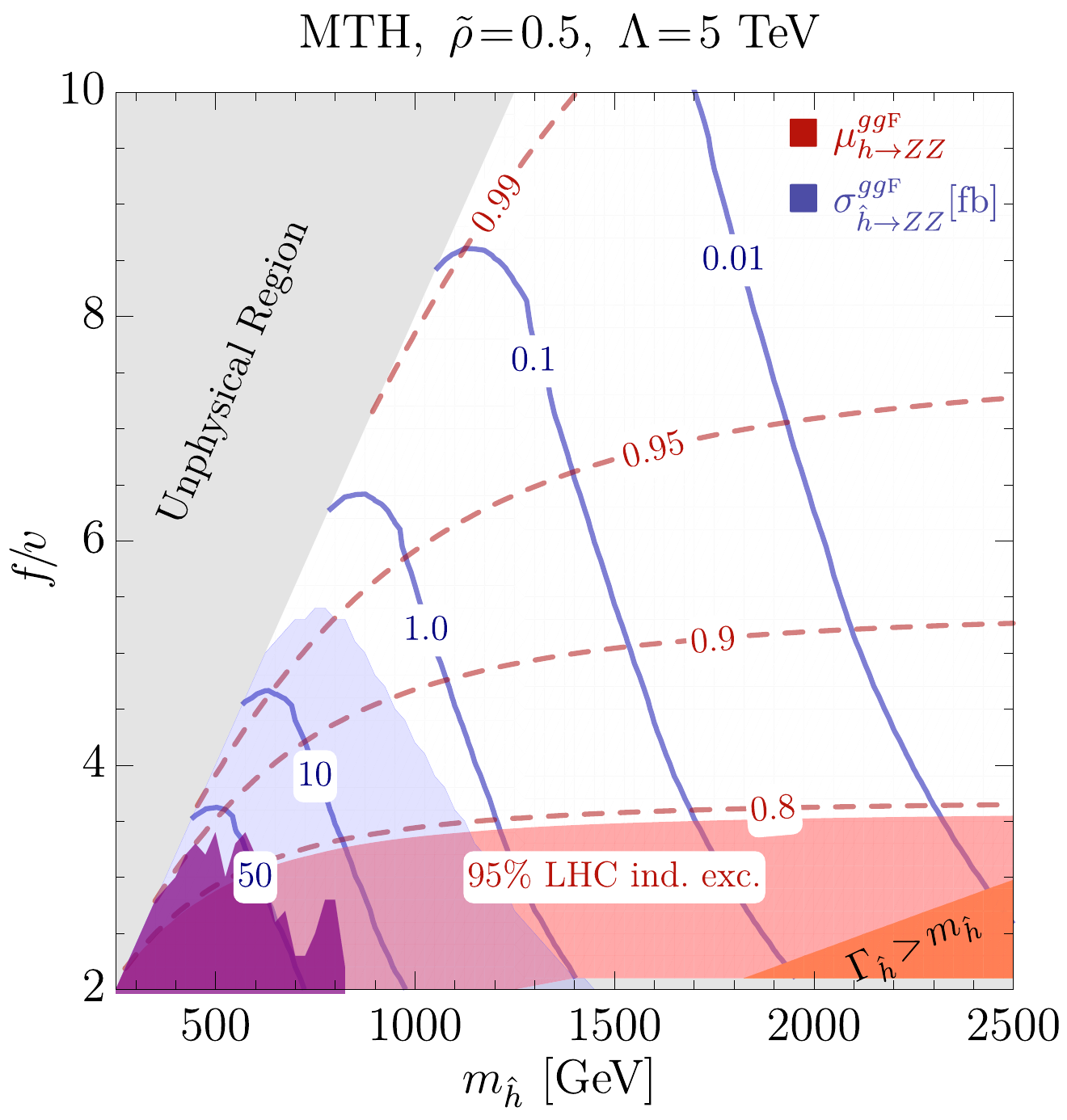}
\caption{These plots show parameter space in $\mht\!-\!f\!/\!v$ plane for the MTH model. The contours of heavy twin Higgs cross-section to the SM $Z$-boson $\sigma^{gg\textsc{f}}_{\hat h\!\to\! ZZ}$~[fb] (solid blue) and SM Higgs signal-strength $\mu^{gg\textsc{f}}_{h\!\to\!ZZ}$ (dashed red) are shown for $\tilde \rho\!=\!0$ (left) and $\tilde \rho\!=\!0.5$ (right). Shaded regions read as: unphysical parameters~(gray), $\Gamma_{\hat h}\!>\!\mht$~(orange), exclusion by the current LHC direct searches $\hat h\!\to\! ZZ$~(purple), the projected reach for direct searches $\hat h\!\to\! ZZ$ at the HL-LHC~(light-blue), and exclusion by the SM Higgs signal-strength measurements at the LHC run-1~(red).}
\label{fig:mthfvvr0}
\end{figure}
The description of shaded regions in Figs.~\ref{fig:mthfvvr0}~and~\ref{fig:mthfhhr0} are as follows: 
\bit\itemsep0em
\item Gray region of Figs.~\ref{fig:mthfvvr0}~and~\ref{fig:mthfhhr0} represents unphysical parameter space excluded by the theoretical constraints as outlined in the previous section.
\item Orange area (Figs.~\ref{fig:mthfvvr0}~and~\ref{fig:mthfhhr0}) captures parameter spaces where heavy twin Higgs width is greater than its mass, $\Gamma_{\hat h}\!>\!\mht$.
\item Red shaded region (Figs.~\ref{fig:mthfvvr0}~and~\ref{fig:mthfhhr0}) is excluded at 95\% C.L. by the combined ATLAS and CMS analysis of the SM Higgs signal-strength ($\mu^{gg\textsc{f}}_{h\!\to\!ZZ}\!<\!0.79$) at the LHC~\cite{Khachatryan:2016vau}.
\item Purple area (Fig.~\ref{fig:mthfvvr0}) shows 95\% C.L. exclusion by direct searches of heavy scalar decaying to the SM $VV$ final state at the ALTAS experiment with 13~TeV~\cite{ATLAS:2017xvp,ATLAS:2017spa,Aaboud:2017itg}.
\item Blue region (Fig.~\ref{fig:mthfvvr0}) presents projected 95\% C.L. exclusion limit for cross-section times branching fraction of heavy twin Higgs to SM vector bosons $\sigma^{gg\textsc{f}}_{\hat h\!\to\!ZZ}$ for the HL-LHC at 14~TeV with an integrated luminosity of 3000~fb$^{-1}$. This estimate is adopted from~\cite{Buttazzo:2015bka}.
\item Green shaded area (Fig.~\ref{fig:mthfhhr0}) shows projected 95\% C.L. exclusion limit for cross-section times branching fraction of heavy Higgs to a pair of SM Higgs bosons $\sigma^{gg\textsc{f}}_{\hat h\!\to\! hh}$ at the HL-LHC, as adopted from~\cite{Buttazzo:2015bka}.
\eit 

From Figs.~\ref{fig:mthfvvr0}~and~\ref{fig:mthfhhr0} one can notice a clear enhancement in the heavy twin Higgs cross-sections to the SM vector bosons and the Higgs boson for $\tilde\rho\!=\!0.5$ as compared to $\tilde \rho\!=\!0$ case. As we discussed above, by turning on the hard breaking effects ($\tilde\rho\!>\!0$) not only cross-sections of heavy twin Higgs to the visible sector would increase due to increase in the mixing angle but also the SM Higgs signal-strength would increase. As a result, on one hand, the observability potential of heavy twin Higgs through direct search would increase at colliders, on the other hand, the indirect constraints from the SM Higgs signal-strength measurements would become stringent. However, we note that for $\tilde \rho\!<\!0$ the mixing angle decreases, hence making it harder to exclude the parameter space via direct or indirect searches at colliders. As discussed above, a generic prediction of the TH models is that heavy Higgs branching fractions to $hh$ is comparable to that of the $VV$, hence the signal of heavy Higgs to a pair of SM Higgs is very important for the discovery potential of TH mechanism. However notice that the HL-LHC reach for $\sigma^{gg\textsc{f}}_{\hat h\!\to\! hh}$ is about an order of magnitude less than that of the $\sigma^{gg\textsc{f}}_{\hat h\!\to\! ZZ}$\,. We summarize results for the MTH parameter scan in Fig.~\ref{fig:mthfvvr0}~and~\ref{fig:mthfhhr0} as: 
\bit\itemsep0em
\item[(i)] The LHC run-1 and early run-2 data only constraints a small region of parameters $f\!/\!v\!\lsim\!3$ for heavy twin Higgs decays to the SM $VV$. 
\item[(ii)] Projected reach of direct searches at the HL-LHC indicates that the MTH model with $\mht\!\lsim\!1\tev$ and $f\!/\!v\!\lsim\!5$ would be accessible. 
\item[(iii)] Direct searches are more sensitive to the low mass region, whereas the indirect constraints are more effective in the high mass regions of the MTH model. Current indirect constraint implies $f\!/\!v\!\lsim\! 3$, however the reach for the signal-strength measurement at the HL-LHC is about $5\%$~\cite{Dawson:2013bba} which implies constraints up to $f\!/\!v\!\lsim\!6$.
\eit
\begin{figure}[t]
\centering
\includegraphics[width=0.49\textwidth]{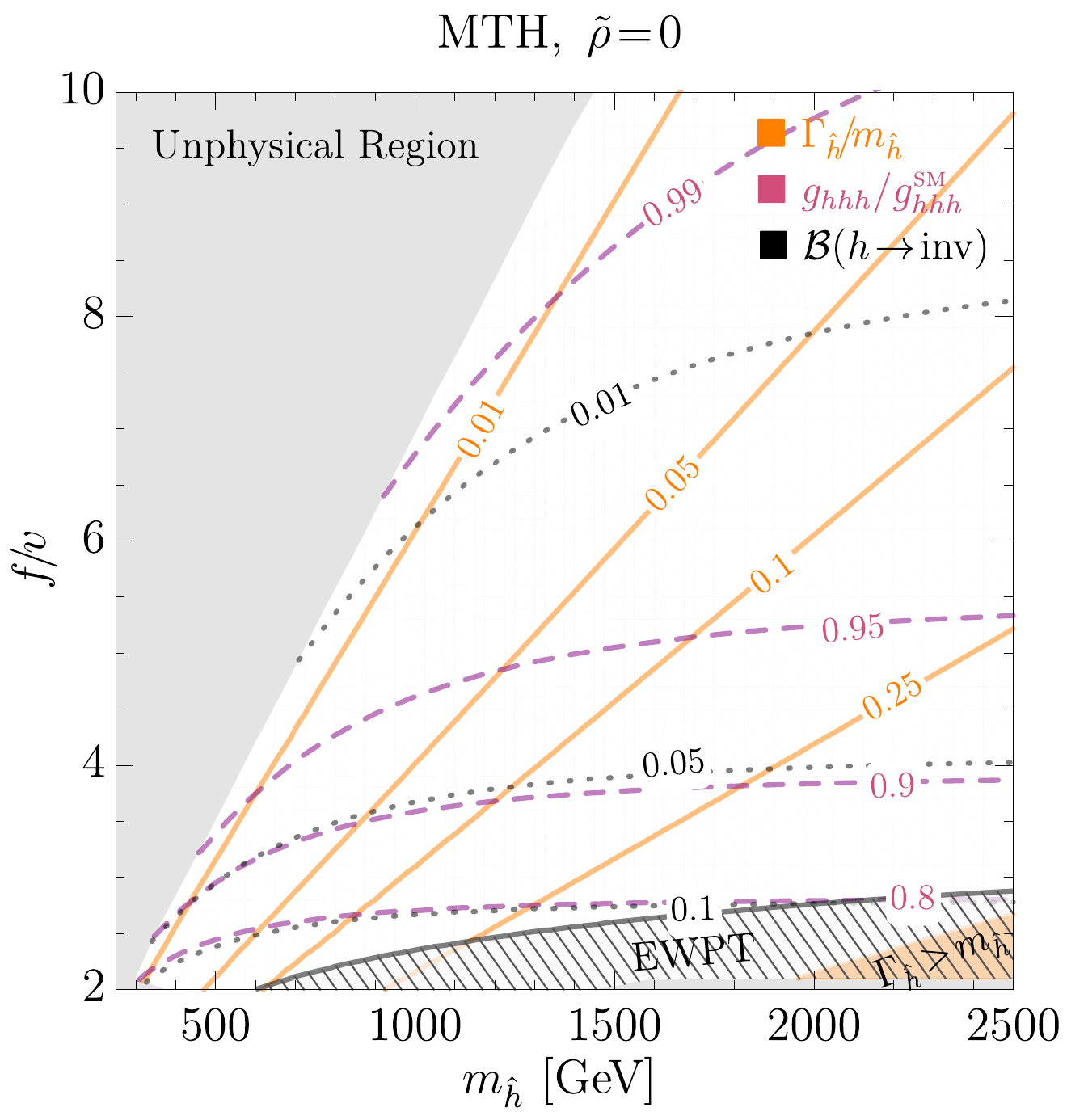}
\includegraphics[width=0.49\textwidth]{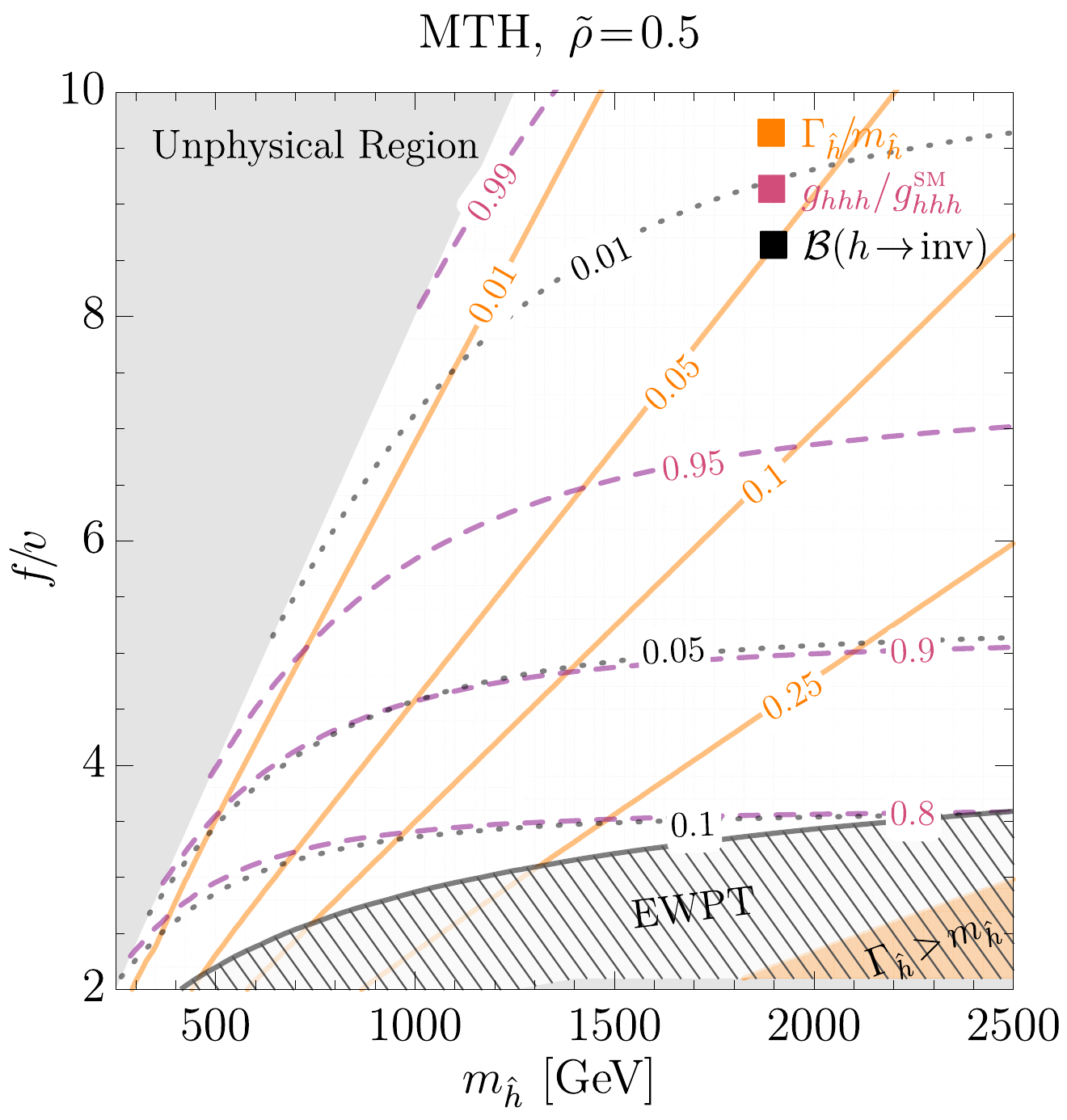}
\caption{These plots show parameter space in $\mht\!-\!f\!/\!v$ plane for the MTH model with $\tilde \rho\!=\!0$ (left) and $\tilde\rho\!=\!0.5$ (right). The contours of the ratio of heavy twin Higgs width over its mass $\Gamma_{\hat h}/\mht$ (solid orange), the ratio of the SM-like Higgs trilinear coupling over its corresponding SM values $g_{hhh}/g^{\textsc{sm}}_{hhh}$ (dashed purple), and the SM Higgs invisible (twin) branching fraction ${\cal B}(h\!\to\! {\rm inv.})$ (dotted). The gray area corresponds to the unphysical parameters, the hatched-gray region is excluded by the EWPT at $95\%$ C.L., and the orange region represents $\Gamma_{\hat h}\!>\!\mht$\,.}
\label{fig:mthinvwidthr0}
\end{figure}

For completeness in Fig.~\ref{fig:mthinvwidthr0} we show the contours of the ratio of heavy twin Higgs width over its mass $\Gamma_{\hat h}/\mht$ (solid orange), the ratio of the SM-like Higgs trilinear coupling over its SM value $g_{hhh}/g^{\textsc{sm}}_{hhh}$ (dashed purple), and the SM Higgs invisible (twin) branching fraction ${\cal B}(h\!\to\! {\rm inv.})$ (dotted), in $\mht\!-\!f\!/\!v$ plane for $\tilde \rho\!=\!0$ (left) and $\tilde\rho\!=\!0.5$ (right). Moreover, in Fig.~\ref{fig:mthinvwidthr0} we also show $95\%$ C.L. constraints on the TH parameter space due to the EWPT (gray hatched region), see Sec.~\ref{ewpt} for details.  Note the indirect probes via the SM-like Higgs trilinear coupling $g_{hhh}/g^{\textsc{sm}}_{hhh}$ and/or the SM Higgs invisible (twin) branching fraction ${\cal B}(h\!\to\! {\rm inv})$ are far less constraining at the LHC~\footnote{The most stringent current bound on the SM Higgs invisible decays ${\cal B}(h\!\to\! {\rm inv.}\!)\!\leq\!0.24$ is by CMS~\cite{Khachatryan:2016whc}.} than the Higgs coupling measurements. However, at the future colliders they can be complimentary probe channels.

\section{Heavy Higgs Phenomenology in Fraternal Twin Higgs} 
\label{Heavy Higgs Phenomenology in the Fraternal Twin Higgs} 

The FTH model has the minimal twin sector matter content required to cancel the radiative corrections to the SM Higgs mass~\cite{Craig:2015pha}. The Lagrangian for the FTH model is,
\beq
\cL_\fth=\cL_{321}+\hat\cL_{\hat3\hat2}-V_{\eff}(H,\widehat H),    \label{lang_mth}
\eeq  
where $\cL_{321}$ is the SM Lagrangian and $V_\eff(H,\widehat H)$ is the scalar potential (same as the MTH model) given in Eq.~\ref{Veff_AB}. However, the FTH model differs from the MTH case in the twin sector Lagrangian, $\hat \cL_{\hat3\hat2}$, especially in the matter content. The most salient features of the FTH model as compared to the MTH case are as follows~\cite{Craig:2015pha}:
\bit\itemsep0em
\item[$\diamond$] Gauge symmetry of the FTH Lagrangian $\hat \cL_{\hat3\hat2}$ is $\widehat{SU}(3)_{c}\!\times\!\widehat{SU}(2)_{L}$, \ie\ there is no twin hypercharge $U(1)_{\hat Y}$ gauge group and hence, no twin photon in the FTH model. 
\item[$\diamond$] Twin fermionic part of the FTH model contains only third generation of twin quarks ($\hat t,\hat b$) and leptons ($\hat \tau,\hat\nu_{\tau}$), which are necessary to cancel the SM radiative corrections to the SM Higgs mass. 
\item[$\diamond$] Twin sector gauge couplings are required to be very close (better than 1\%) to that of the SM at the cutoff scale $\Lambda$, \ie\
\beq
\hat g_s(\Lambda)\simeq g_s(\Lambda), \lsp \hat g(\Lambda)\simeq g(\Lambda).
\eeq
Similarly, twin top Yukawa coupling $\hat y_{\hat t}$ is required to be very close to the SM top Yukawa coupling, \ie\ $\hat y_{\hat t}\simeq y_t$. These requirements are motivated by the naturalness arguments in cancellation of the radiative corrections. 
\item[$\diamond$] Contrary to the top Yukawa coupling, the  light twin fermions Yukawa couplings ($\hat y_{\hat b},\hat y_{\hat \tau}$) could be substantially different from their SM  counterparts. 
\item[$\diamond$] Twin sector masses of the weak gauge bosons $m_{\hat V}$ and fermions $m_{\hat f}$ scale w.r.t. their SM counterparts as, 
\beq
m_{\hat V} =\frac{\hat v}{v}m_{V}\,,    \lsp m_{\hat f} =\frac{\hat v}{v}\frac{\hat y_{\hat f}}{y_f}m_{f}~,
\eeq
where the light twin fermions receive an extra $\hat y_{\hat f}/y_f$ enhancement in their masses (and also in their couplings) as compared to the MTH model.
\item[$\diamond$] The light twin fermions Yukawa couplings ($\hat y_{\hat b},\hat y_{\hat \tau}$) are essentially free parameters. Which implies an enhancement of the heavy Higgs (and the SM Higgs if $m_{\hat f}\!<\!m_h/2$) decay rates to light twin fermions by a factor $y_{\hat f}^2/y_f^2$ as compared to the MTH case.
\item[$\diamond$] The most remarkable feature of the FTH model is the relative large QCD confinement scale (due to faster running of the twin QCD coupling constant $\hat\alpha_{\textsc{qcd}}$) and subsequent hadronization of twin gluons/bottom quarks to twin glueball/bottomonium states, respectively. In particular, the $0^{++}$ twin hadrons mix with the Higgs bosons and lead to prompt and/\!or displaced exotic decays to the SM light fermions. 
\eit

In the FTH model, the heavy twin Higgs decay rates and cross-sections to the SM sector would be almost same as for the MTH case (Sec.~\ref{Heavy Higgs Phenomenology in the Mirror Twin Higgs}). Moreover, the twin weak gauge bosons $\hat W,\hat Z$ and twin top $\hat t$ of the FTH have the same masses and coupling to heavy twin Higgs as those for the MTH case. Therefore, their the branching fractions/cross-sections would not change substantially as compared to the MTH case. However, light twin fermion Yukawa couplings can be enhanced, as a result their production cross-sections would get enhancement. To be precise, heavy twin Higgs decay rates to the light twin fermions get an extra $(\hat y_{\hat f}/y_f)^2$ enhancement factor as compared to that of the MTH model. The on-shell decays of heavy Higgs to a pair of twin massive vectors bosons ($\hat h\to \hat V\hat V$) remain same as in the MTH scenario. However, the off-shell decays $\hat h\to \hat V^\ast\hat V^\ast\to 4\hat f$ involve total widths of the vector bosons $\Gamma_{\hat V}$, which change as compared to the MTH case due to fewer light twin fermions. In the FTH model, total decay widths of the twin bosons $\hat W$ and $\hat Z$ are (assuming $m_{\hat f}\ll m_{\hat V}$),
\beq
\Gamma(\hat W\to \hat \tau \bar{\hat\nu}_\tau)=\frac{\hat g^2 m_{\hat W}}{48\pi},     \lsp \Gamma(\hat Z\to \hat f \bar{\hat f})=\frac{\hat g^2 m_{\hat Z}}{48\pi }\big(\hat g_V^2+\hat g_A^2\big),
\eeq
where $\hat g_V$ and $\hat g_A$ are the vector and axial couplings of the twin fermions $\hat f=(\hat b,\hat \tau,\hat \nu_\tau)$. 

\begin{figure}[t]
\centering
\includegraphics[width=0.49\textwidth]{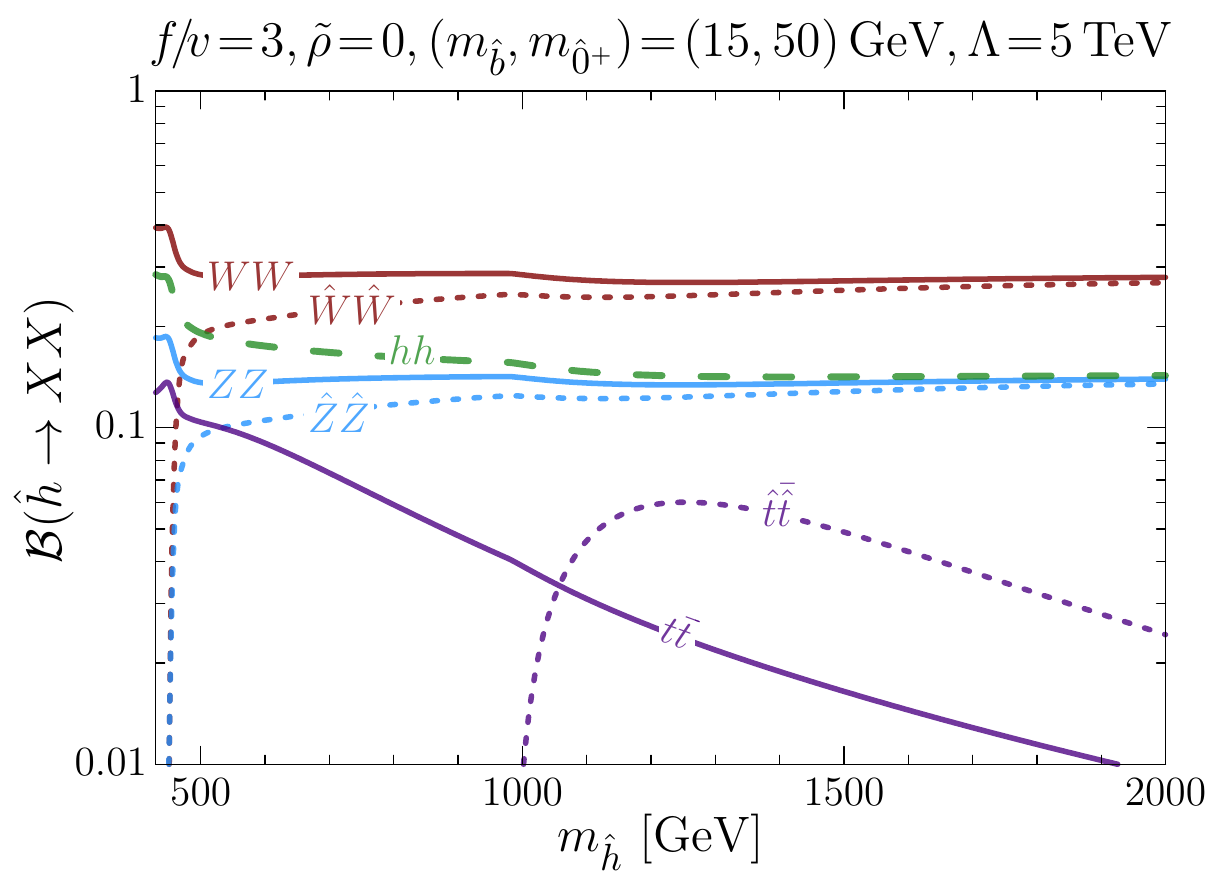}\!
\includegraphics[width=0.49\textwidth]{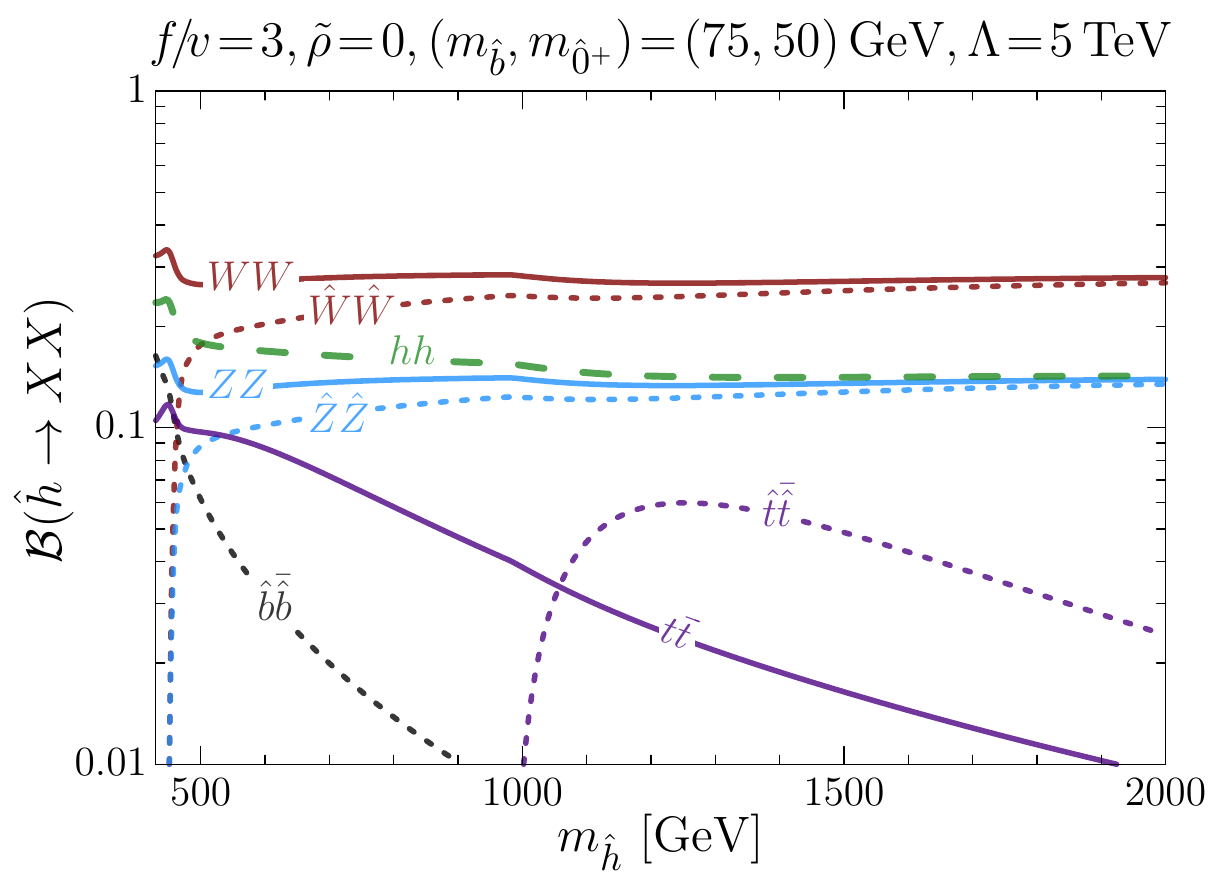}
\caption{These plots show branching ratios of heavy twin Higgs to different final states as a function its mass in the FTH model. The left and right plots are for $(m_{\hat b},m_{\hat 0^+})\!=\! (15,50)\gev$ and $(75,50)\gev$, respectively, where $f\!/\!v\!=\!3$ and $\tilde \rho\!=\!0$.}
\label{fig:fthbrs}
\end{figure}
In Fig.~\ref{fig:fthbrs}, we show the heavy twin Higgs branching fractions ($\geq\!1\%$) to the SM and twin states, where the left- and right-panel are for $y_{\hat b}\!/\!y_b\!\simeq\!1.2$ and 6, respectively, for fix values of $f\!/\!v\!=\!3$, $\tilde \rho\!=\!0$, and $\hat \Lambda_{\textsc{qcd}}\!=\!7.35\gev$ (which corresponds to the lightest glueball mass $m_{\hat 0^+}\!=\!50\gev$). For simplicity, in this work we take twin tau Yukawa as $y_{\hat \tau}\!/\!y_\tau\!=\!y_{\hat b}\!/\!y_b$, which fixes the twin sector fermion masses up to only one free parameter, the twin bottom Yukawa coupling $\hat y_{\hat b}$ (or $m_{\hat b}$). Notice the heavy twin Higgs branching fractions do not substantially change for the twin massive gauge bosons and twin top for $y_{\hat b}\!/\!y_b\!\simeq\!1.2$ and 6 cases. However, the heavy twin Higgs branching fraction for twin bottom quark, ${\cal B}(\hat h\!\to\!\hat b\hat b)$ (shown as dotted curve in Fig.~\ref{fig:fthbrs}), increases about an order of magnitude for a given $\mht$ value as we increase $y_{\hat b}\!/\!y_b$ from $\sim\!1$ to 6. The effect of explicit hard breaking parameter $\tilde\rho$ on the heavy Higgs branching fractions is similar to the MTH case, i.e. there is an enhancement in the SM branching fractions and subsequently suppression to twin sector branching ratios for $\tilde\rho\!>\!0$, and vice versa for $\tilde\rho\!<\!0$. 

The most notable difference of the FTH model as compared to the MTH case, is the running of twin QCD coupling constant $\hat \alpha_{\textsc{qcd}}$ from the cutoff scale $\Lambda$ to the confinement scale $\hat \Lambda_{\textsc{qcd}}$. Since the number of quark flavors is less in the FTH (so faster running) as a result the confinement scale $\hat \Lambda_{\textsc{qcd}}$ would be larger than the SM QCD confinement scale. In the following numerical analysis we employ the NNLO twin QCD corrected $\hat \alpha_{\textsc{qcd}}$, however, we require at the cutoff scale~$\Lambda$,\hfill~
\begin{wrapfigure}{r}{0.5\textwidth}
\centering
\includegraphics[width=0.5\textwidth]{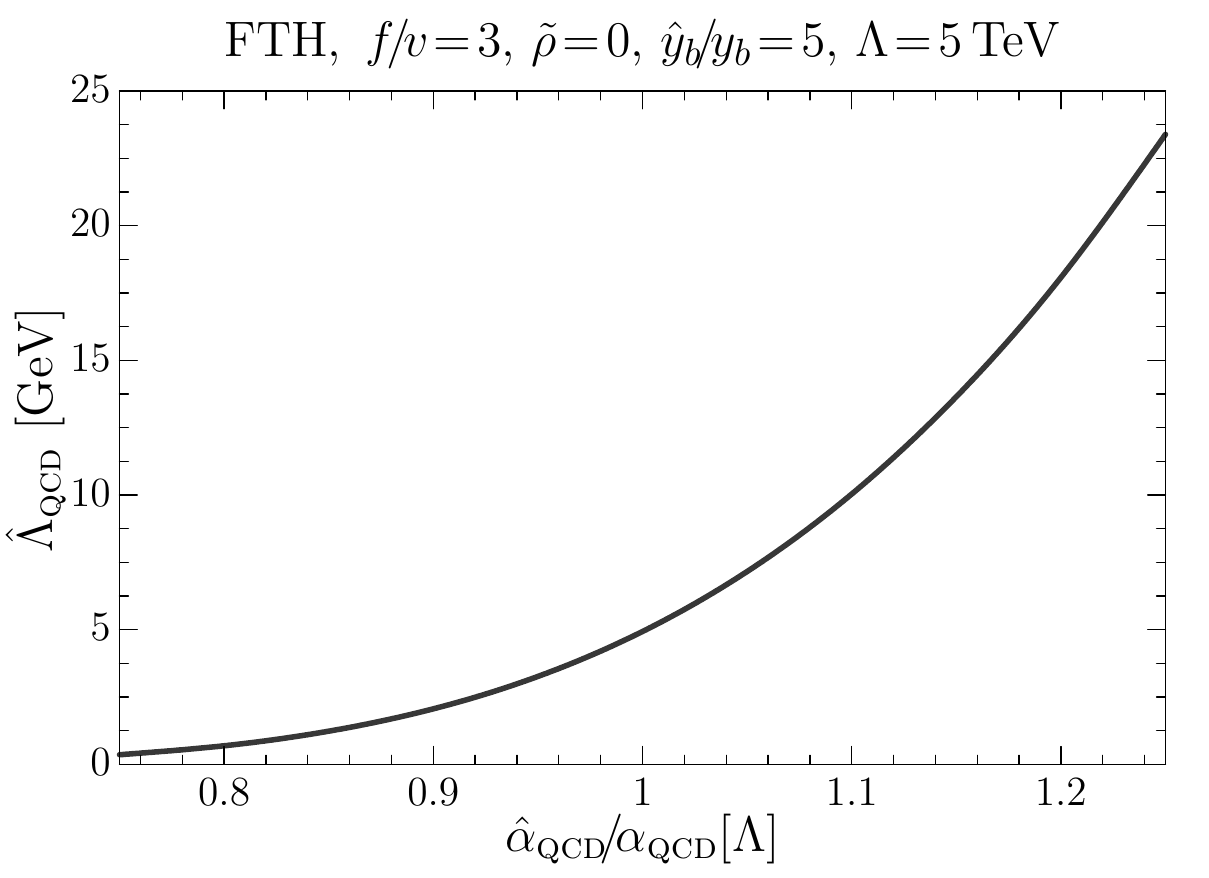}
\caption{The twin confinement scale $\hat \Lambda_{\textsc{qcd}}$ as a function of $\hat \alpha_{\textsc{qcd}}\!/\!\alpha_{\textsc{qcd}}$ at the cutoff $\Lambda\!=\!5\tev$.}
\label{fig:alstlam}
\end{wrapfigure}
\beq
0.75 \lsim \frac{\hat \alpha_{\textsc{qcd}}(\Lambda)}{\alpha_{\textsc{qcd}}(\Lambda)}\lsim 1.25.
\eeq
The above variation in the twin strong coupling constant induce a rather mild fine-tuning $\sim\!30\%$ in the SM Higgs mass. A small variation of $\hat \alpha_{\textsc{qcd}}$ as compared to $\alpha_{\textsc{qcd}}$ at the high scale $\Lambda$ significantly change the confinement scale which has very crucial phenomenological consequences. In Fig.~\ref{fig:alstlam} we plot the twin confinement scale $\hat \Lambda_{\textsc{qcd}}$ as function of $\hat \alpha_{\textsc{qcd}}\!/\!\alpha_{\textsc{qcd}}$ at the cutoff scale $\Lambda\!=\!5\tev$, where we kept fix $f\!/\!v\!=\!3$, $\tilde \rho\!=\!0$, and $y_{\hat b}/y_b\!=\!5$.

\subsection{Fraternal Twin Hadron Production and Decays}
In the FTH model due to fewer twin quark flavors and potentially large confinement scale $\hat \Lambda_{\textsc{qcd}}$, the twin QCD confines to form twin glueball and bottomonium $[\hat b \bar{\hat b}]$ states. The lightest of the twin glueballs is a $0^{++}$ state $\hat G_{0^+}$ with mass $m_{\hat 0^+}\!\!\simeq\!6.8\hat \Lambda_{\textsc{qcd}}$. Notice that the lightest glueball $\hat G_{0^+}$ has the correct quantum numbers to mix with the SM Higgs (and also twin Higgs) which allows its decay to the SM light fermions. In analogy to the SM bottomonium spectra, the lightest twin bottomonium state is $\hat \eta (0^{-+})$, and next to lightest is $\hat \Upsilon (1^{--})$ state. However, the $0^{++}$ p-wave state $\hat \chi_0$ with mass $m_{\hat \chi}\simeq 2m_{\hat b}$ has the correct quantum numbers to mix with the TH scalar sector such that $\hat \chi_0$ decays in the visible sector are also possible. Phenomenological aspects of production and decays of the twin glueball and twin bottomonium states due to the SM-like Higgs has been discussed in~\cite{Craig:2015pha,Curtin:2015fna,Csaki:2015fba,Chacko:2015fbc,Pierce:2017taw}, see also~\cite{Strassler:2006im,Strassler:2006ri,
Han:2007ae,Kang:2008ea,Juknevich:2009ji,Juknevich:2009gg,Curtin:2013fra,Cheng:2015buv,Cheng:2016uqk,Buchmueller:2017uqu}. In this work, we follow these analyses closely but focus on the production of twin hadrons due to heavy twin Higgs.

Since twin hadronization is poorly understood, especially, without the proper knowledge of the non-perturbative twin QCD effects, its hard to make a precise estimate of twin hadron production. However, the perturbative estimates can be made quite accurately and the ignorance of twin hadronization can be controllably parametrized, see for instance~\cite{Craig:2015pha,Juknevich:2009gg,Curtin:2015fna,Chacko:2015fbc,Pierce:2017taw}. The twin QCD hadronization of twin gluons and bottom quarks lead to a dozen of meta-stable states in their spectrums. However, we are interested in the production of $0^{++}$ states, i.e. twin glueball $\hat G_{0^+}$ and twin bottomonium $\hat \chi_0$, via the scalar sector of the FTH model. Such that the $0^{++}$ twin hadrons mix with the SM Higgs and lead to exotic decays to the SM light fermions. 

We parametrize heavy twin Higgs partial decay rate to twin glueball $\hat G_{0^+}$ and twin bottomonium $\hat \chi_0$ states as,
\begin{align}
\Gamma(\hat h\to \hat G_{0^+})&=\Gamma(\hat h\to \hat g\hat g)\, \langle N_{\hat G}\rangle\, \hat\kappa_0 \,\sqrt{1-\tfrac{4m_{\hat 0^+}^2}{m_{\hat h}^2}}\,,    \\
\Gamma(\hat h\to \hat \chi_0)&=\Gamma(\hat h\to \hat b\hat b)\, \langle N_{[\hat b\bar{\hat b}]}\rangle\, \hat\kappa_0 \,\sqrt{1-\tfrac{4m_{\hat \chi}^2}{m_{\hat h}^2}}\,,
\end{align}
where $\langle N_{\hat G}\rangle$ and $\langle N_{[\hat b\bar{\hat b}]}\rangle$ are the hadronic multiplicities of glueball and bottomonium states at the heavy Higgs mass $m_{\hat h}$, respectively. Whereas, $\hat\kappa_0 $ is the fraction of glueball or bottomonium states in the $0^{++}$ state. Twin QCD hadronic average multiplicity $ \langle N_{\rm had}\rangle$ is defined as~\cite{Ellis:1991qj}, 
\begin{align}
\langle N_{\rm had}(s)\rangle = N_0\, \exp\!\bigg[\frac{1}{\hat b_0}\sqrt{\frac{6}{\pi \hat \alpha_{\textsc{qcd}}(s)}}+\left(\frac14+\frac{5 n_{\hat f}}{54\pi \hat b_0}\right)\ln \hat\alpha_{\textsc{qcd}}(s)\bigg],
\end{align}
where $\hat b_0 \!=\!(33-2n_{\hat f})/(12\pi)$ is the twin QCD beta function, $\hat\alpha_{\textsc{qcd}}(s)$ is the twin QCD coupling evaluated at the center-of-mass energy $\sqrt s$, and $n_{\hat f}$ are the number of twin quark flavors below the confinement scale $\hat \Lambda_{\textsc{qcd}}$. The normalization constant $N_0$ is a free parameter which we fix by setting the number of twin hadrons at a fixed low energy. For example, for a glueball of mass $m_{\hat 0^+}\!=\!50 \gev$, a scalar at mass $125\gev$ can produce roughly two glueballs, so we fix $N_0$ by the condition $\langle N_{\hat G}\rangle_{125}\!=\!2$. The above hadron multiplicity for the dark QCD has been tested by~\cite{Schwaller:2015gea} with showering/hadronization in {\tt PYTHIA}~\cite{Sjostrand:2014zea} and a very good agreement has been found.

\begin{figure}[t]
\centering
\includegraphics[width=0.49\textwidth]{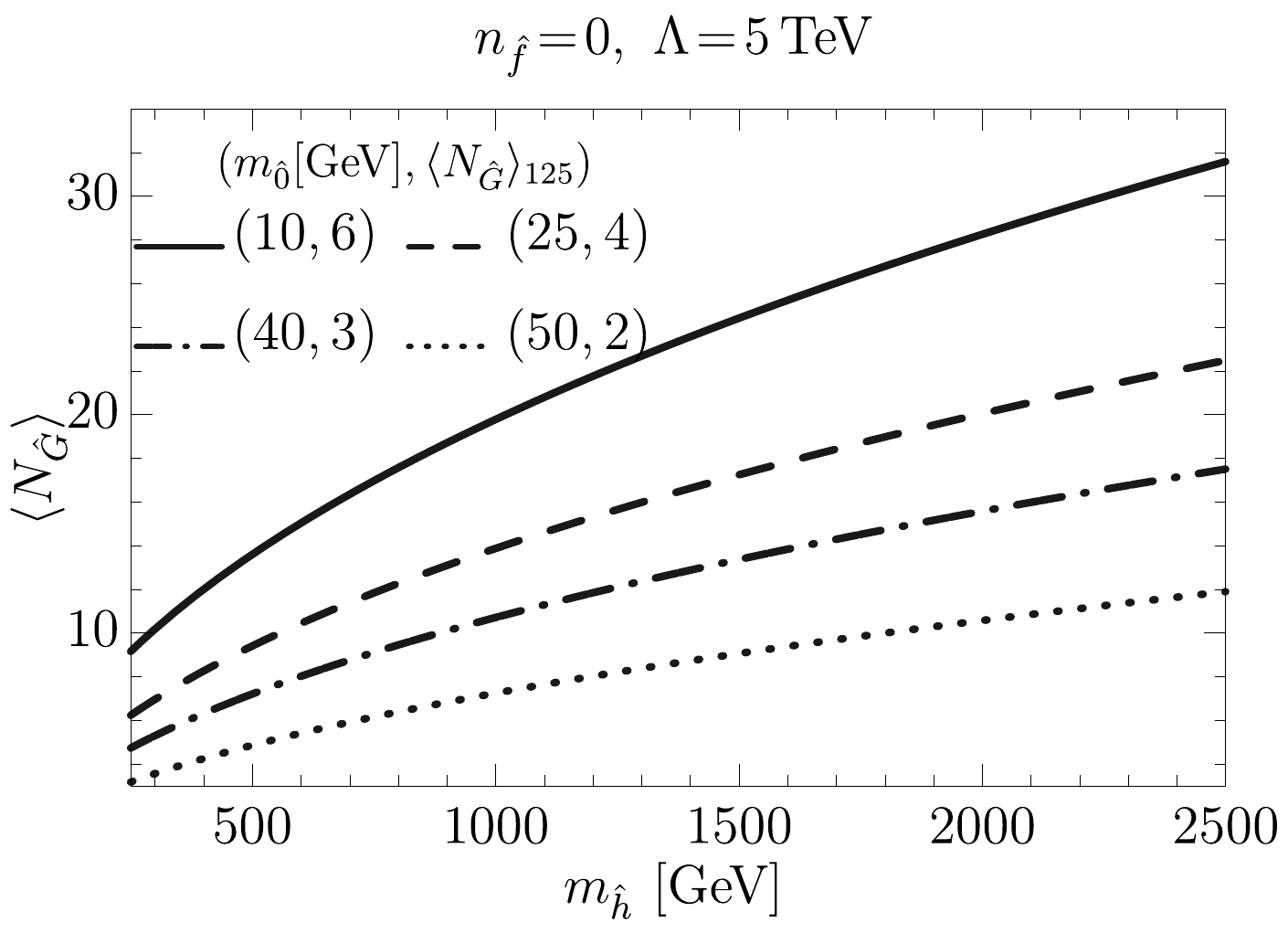}
\includegraphics[width=0.49\textwidth]{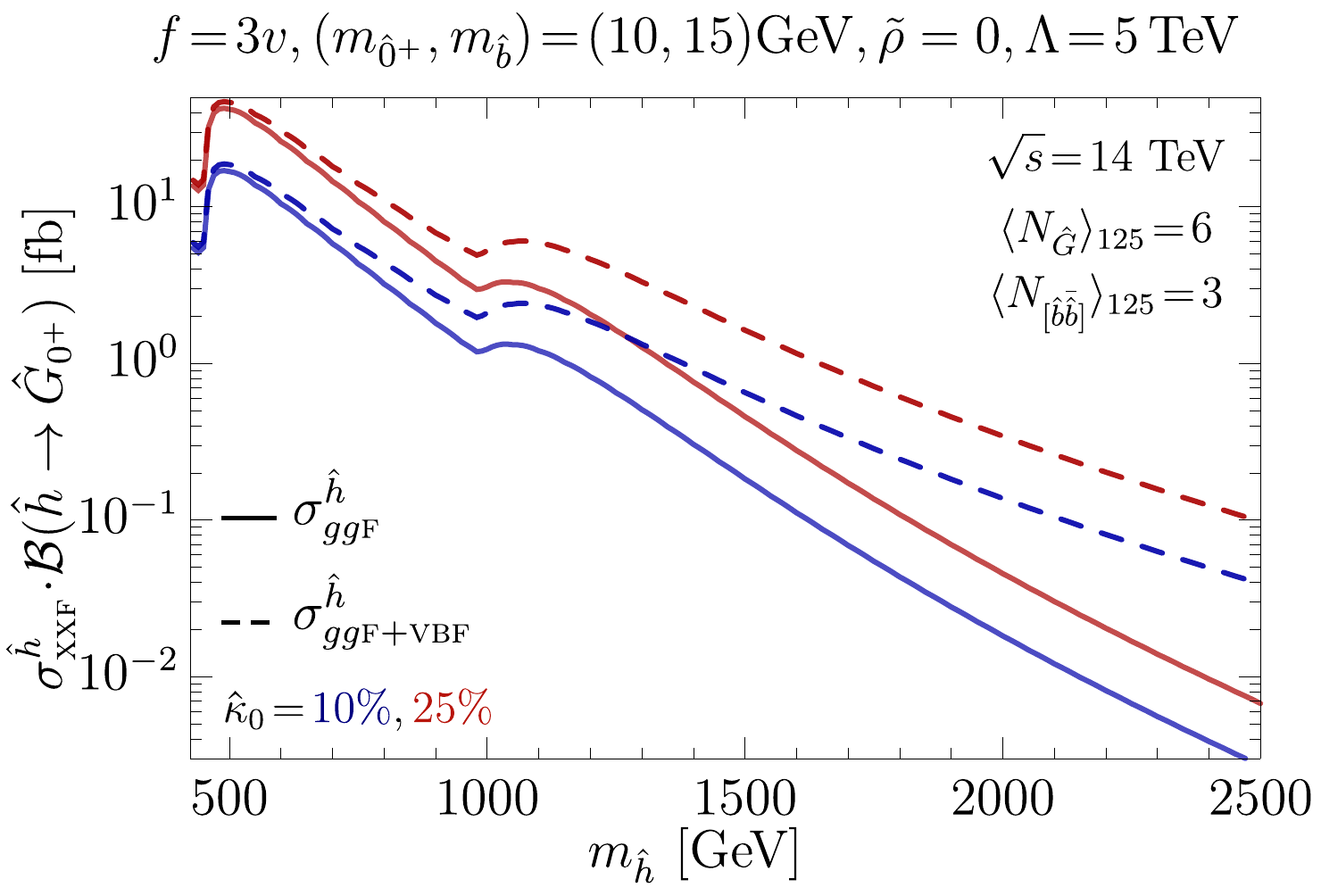}
\caption{The left-graph shows average glueball multiplicity $\langle N_{\hat G}\rangle$ as a function of the heavy Higgs mass for different choices of the lightest glueball mass and the corresponding normalized number of glueball states at the SM Higgs mass. In the right-panel we plot cross-section times branching fraction of the heavy twin Higgs decaying to the lightest glueballs $\hat G_{0^+}$ with mass $m_{\hat 0^+}\!=\!10\gev$ as a function of the twin Higgs mass. The twin Higgs at the $14\tev$ LHC is produced via gg{\textsc f} (solid) and $gg{\textsc f}\!+\!{\textsc{vbf}}$ (dashed). We take $\hat\kappa_0 \!=\!10\%$ (blue) and $\hat\kappa_0 \!=\!25\%$ (red) for the lightest glueball multiplicity.}
\label{fig:multi_xsecgb}
\end{figure}

In Fig.~\ref{fig:multi_xsecgb} (left-panel) we plot the average multiplicity of twin glueballs as a function of the twin Higgs mass for different choices of the lightest glueball mass $m_{\hat 0^+}$. Note that we normalize the glueball multiplicity w.r.t. to the SM Higgs, \ie\ we take $\langle N_{\hat G}\rangle_{125}\!=\!(6,4,3,2)$ corresponding to $m_{\hat0^+}\!=\!(10,25,40,50)\gev$, respectively, which are rather conservative numbers. Moreover, when the twin bottomonium states are heavier than the lightest glueball, then through the relaxation/annihilation these bottomonium states decay to twin gluons which then hadronize to form the glueballs. In the following we assume if $m_{[\hat b\bar{\hat b}]}\!>\!2m_{\hat0^+}$ then all the bottomonium states decay to the glueballs, \ie\ $[\hat b\bar{\hat b}]\to{\rm glueballs}$. The right-panel of Fig.~\ref{fig:multi_xsecgb} show the production cross-section of the lightest glueball state with mass $m_{\hat 0^+}\!=\!10\gev$ via the heavy twin Higgs. The heavy Higgs production is considered by the gg{\textsc f} (solid curve) and gg{\textsc f}+{\sc vbf} (dashed curve) at the LHC with center-of-mass energy $14\tev$. As mentioned above, around a dozen (depending on their masses) meta-stable glueball states are formed and the probability to produce the lightest glueball $\hat G_{0^+}$ is parameterized by $\hat\kappa_0 $, we take a conservative choice of $\hat\kappa_0 \!=\!10\%$ (blue curves) and $\hat\kappa_0 \!=\!25\%$ (red curves). In Fig.~\ref{fig:multi_xsecgb} (right-panel) we consider $m_{\hat b}\!=\!15\gev$ ($n_{\hat f}\!=\!0$), such that only stable twin hadrons are the glueballs, whereas, other parameters are $f/v\!=\!3$ and $\tilde\rho\!=\!0$. The bumps in the curves of glueball cross-section indicate thresholds for the twin $\hat Z$ and top quark $\hat t$, as the decays of twin $\hat Z$ boson and $\hat t$ quark to twin bottom $\hat b$ quarks lead to hadronize and decay to the glueball states. Note that there could be non-perturbative effects in the production of the twin glueballs as large as $10\%$ as argued in~\cite{Craig:2015pha}, however, we assume they are within the uncertainty of our approximation for the average hadron multiplicity and the $\hat\kappa_0 $ parameter. Nevertheless, the above parameterization provides a reasonable estimate for the production of fraternal twin hadrons. 

\begin{figure}[t]
\centering
\includegraphics[width=0.7\textwidth]{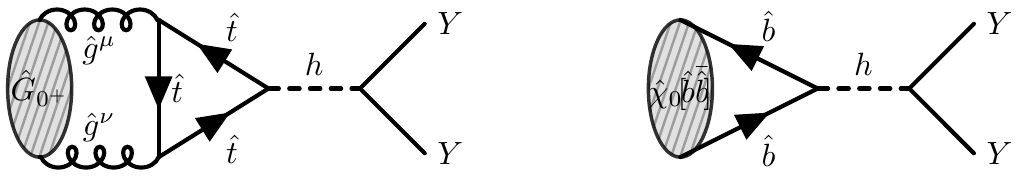}
\caption{Schematic diagrams of twin glueball $\hat G_{0^+}$ and twin bottomonium $\hat \chi_0$ decays to the SM states $Y$ (mostly bottom quarks) via an off-shell SM Higgs $h$.}
\label{fig:twinggbbh}
\end{figure}

The lightest twin glueball $\hat G_{0^+}$ mixes with the TH scalar sector,  which leads to its decay to the SM light quarks and leptons, see Fig.~\ref{fig:twinggbbh}~(left). In the following, we calculate decay width of the twin glueball $\hat G_{0^+}$ to the SM light fermions. The first step is to calculate the twin gluon coupling to the SM Higgs~\footnote{Note that the twin $0^{++}$ hadrons also decay to the SM light fermions via an off-shell heavy twin Higgs~$\hat h$, however, these processes can be neglected as they are suppressed by $(m_h/\mht)^4$ in comparison to the SM Higgs mediated decays.}, which is given by, 
\beq
\cL^{\rm int}_{h\hat g\hat g}= c_{\hat g\hat g}\,\hat g_{h}
{\rm Tr}\big[G^a_{\mu\nu}G^{a,\mu\nu}\big]\,h\,,\lsp{\rm where}\hsp c_{\hat g\hat g}\equiv-\frac{\hat \alpha_{\textsc{qcd}}}{4\pi \hat v} \sum_{i}\hat F_{1/2}(\tau_{i}),
\eeq
where $i\!=\!\hat t,\hat b$ (dominant contribution is by twin top loop) and in the limit $\tau_{\hat t}\!=\!4m_{\hat t}^2/m_h^2\gg1$ the triangle loop function $\hat F_{1/2}(\tau_{\hat t})\!\to\! -4/3$, whereas $\hat g_h\!=\!-\sin\!\alpha$ (see also Fig.~\ref{fig:couplings}). Hence in the low-energy effective theory, twin gluon coupling to the SM Higgs is
\beq
\cL^{\rm int}_{h\hat g\hat g}= -\frac{\hat \alpha_{\textsc{qcd}}}{3\pi \hat v} \sin\!\alpha\,
{\rm Tr}\big[G^a_{\mu\nu}G^{a,\mu\nu}\big]\,h\,.	\label{twinggh}
\eeq
Now following~\cite{Juknevich:2009gg} we can write decay rate of the glueball to the SM light states $Y$ as, 
\beq
\Gamma(\hat G_{0^+}\to YY)=\bigg(\frac{\hat \alpha_{\textsc{qcd}} \sin\!2\alpha ~{\bf F}_{\hat 0^+}}{6\pi\, \hat v\,(m_h^2-m_{\hat 0^+}^2)}\bigg)^2~\Gamma^{\rm SM}(h\to YY)\big\vert_{m_h\!=\!m_{\hat 0^+}}~,
\eeq
where ${\bf F}_{\hat 0^+}\!\equiv\!\langle 0|{\rm Tr}[\hat G^a_{\mu\nu}\hat G^{a,\mu\nu}]|\hat 0^{++}\rangle$ is the decay constant of the lightest glueball $\hat G_{0^+}$ and its value by lattice is $4\pi \hat \alpha_{\textsc{qcd}}{\bf F}_{\hat 0^+}\!=\!3.06 \,m_{\hat 0^+}^3$~\cite{Lucini:2004my}. Above, $\Gamma^{\rm SM}(h\to YY)$ is the SM Higgs decay rate to the SM light states $Y$ evaluated at the twin glueball mass $m_{\hat 0^+}$.

Assuming the twin leptons $(\hat \ell\!=\!\hat \tau,\hat\nu)$ are heavy enough such that the invisible twin weak decays of $\hat \chi_0\to\hat \Upsilon \hat\ell\hat\ell$ are kinematically forbidden. Then, as mentioned above, for $m_{\hat \chi}\!>\!2m_{\hat 0^+}$, bottomonium state $\hat \chi_0$ decays to the glueballs, otherwise for $m_{\hat \chi}\!<\!2m_{\hat 0^+}$, $\hat \chi_0$ decays to the visible sector fermions. The decays of $\hat \chi_0$ state to the SM sector would meditate through the SM Higgs, as shown in Fig.~\ref{fig:twinggbbh} (right), and the interaction is given by, 
\beq
\cL^{\rm int}_{h\hat b\hat b}= -\frac{y_{\hat b}}{\sqrt 2} \hat g_h\, h\hat b\bar {\hat b}=\frac{m_{\hat b}}{\hat v} \sin\!\alpha\,h\,\hat b\bar {\hat b}\,.    \label{twinbbh}
\eeq
An estimate of the bottomonium state $\hat \chi_0$ decay to the SM light fermions via an off-shell Higgs using non-relativistic quantum mechanism is~\cite{Craig:2015pha},
\beq
\Gamma(\hat \chi_{0}\to YY)=\frac{27}{4\pi}\big\vert R^\prime(0)\big\vert^2\,\frac{\sin^2(2\alpha) \,m_{\hat\chi}^2}{32\hat v^2 \,(m_h^2-m_{\hat\chi}^2)^2}\,\Gamma^{\rm SM}(h\to YY)\big\vert_{m_h\!=\!m_{\hat \chi}}~,
\eeq
where $ R^\prime(0)$ is derivative of the radial wave-function of p-wave state $\hat \chi_0$ w.r.t. its radius. We use the estimate of $|R^\prime(0)|^2$ given by~\cite{Craig:2015pha}, which employs linear confining potential, but neglecting the relativistic corrections which could be potentially important. The estimate of $|R^\prime(0)|^2$ for $\hat \chi_0$ state is~\cite{Craig:2015pha}, 
\beq
\big\vert R^\prime(0)\big\vert^2\approx 0.002~\frac{m_{\hat \chi}^{5/3}m_{\hat 0^+}^{10/3}}{m_h}\,.
\eeq

In the following subsection we calculate the production cross-section for the twin hadrons via heavy twin Higgs in $\mht\!-\!f\!/\!v$ plane. The decay length of the twin $0^{++}$ hadron is, 
\beq
c\tau_{0^{++}}=\frac{\hbar c}{\Gamma_{0^{++}}}.
\eeq
Depending on kinematics of the produced twin hadrons ($0^{++}$) their decay-length $c\tau_{0^{++}}$ can be from very short $\lsim 10^{-3}\,{\rm [m]}$ to very long $\gsim 10^2\,{\rm [m]}$. For very short decay-length the daughter decays to the SM light fermions would be prompt, so the signature at the colliders could be high multiplicity jets, see \eg~\cite{Knapen:2016hky,Cohen:2017pzm}. On the contrary, for very long decay-length the twin-hadrons would escape the detectors and signatures would be missing energy.  However, the most promising signals would appear for the case when the twin hadrons decay-length is of the size of detectors, i.e. $10^{-3}\,{\rm m}\lsim c\tau_{0^{++}}\lsim 10^2\,{\rm m}$, such that the signatures would appear as displaced vertices at the colliders. There are ongoing searches at the LHC for long-lived particles at the ATLAS~\cite{ATLAS:2016olj,Aad:2015rba,Aad:2015uaa}, the CMS~\cite{CMS:2014wda,CMS:2014hka} and the LHCb~\cite{Aaij:2016isa} experiments.

\subsection{Fraternal Twin Hadron Phenomenology}
In this subsection, we calculate the production cross-section of twin glueball $\hat G_{0^+}$ and twin bottomonium $\hat \chi_0$ states via heavy twin Higgs at the $14\tev$ LHC~\footnote{In this work, we do not discuss the production of twin hadrons via the SM Higgs in the FTH model, however, we refer to Refs.~\cite{Curtin:2015fna,Csaki:2015fba,Chacko:2015fbc,Pierce:2017taw}.}. For this purpose, we choose several benchmark cases where either the lightest twin hadron is a glueball or a bottomonium state. 

\begin{figure}[t]
\centering
\includegraphics[width=0.49\textwidth]{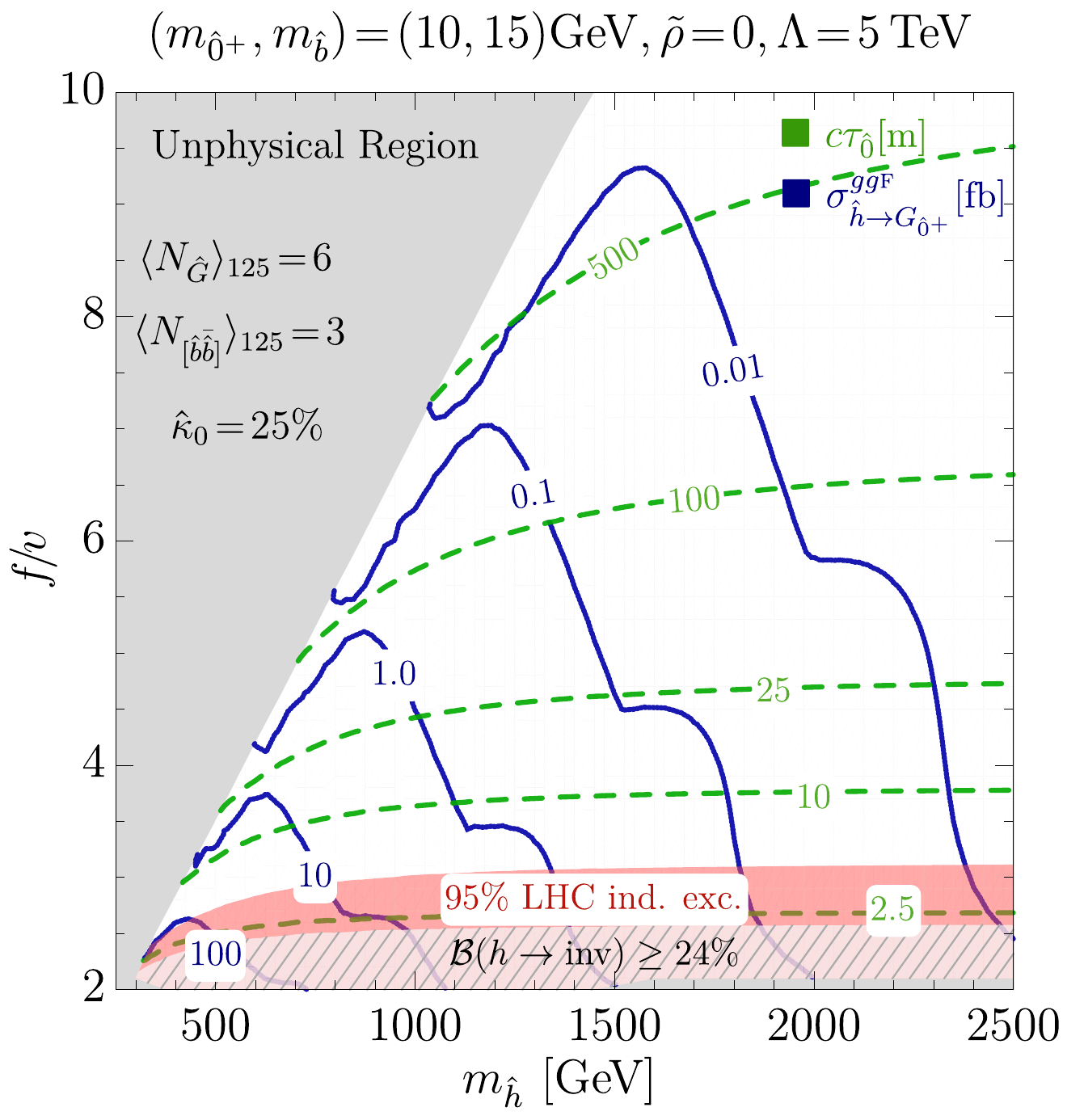}
\includegraphics[width=0.49\textwidth]{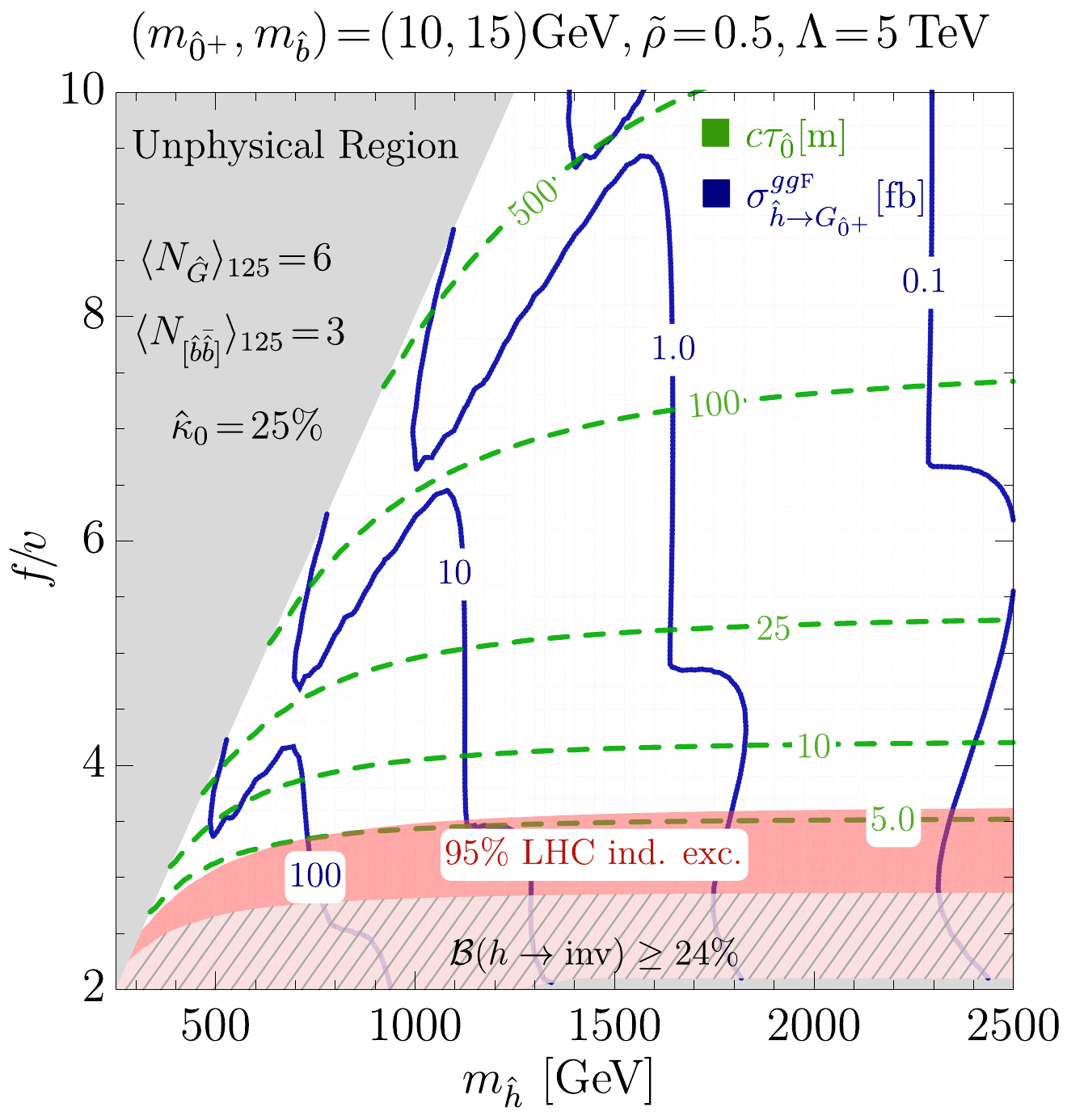}
\caption{These plots show parameter space in $\mht\!-\!f\!/\!v$ plane for the FTH model. The contours of heavy twin Higgs cross-section to twin glueball $\hat G_{0^+}$ with mass $m_{\hat 0^+}\!=\!10\gev$ (solid blue) and the  glueball decay-length $c\tau_{\hat 0}$ (dashed green) are shown for $\tilde \rho\!=\!0$ (left) and $\tilde \rho\!=\!0.5$ (right).}
\label{fig:mthfovrmbt15mgb10}
\end{figure}

In Fig.~\ref{fig:mthfovrmbt15mgb10} we plot the contours of production cross-section~[fb] (solid blue) and proper decay-length~[m] (dashed green) of glueballs $\hat G_{0^+}$ with mass $m_{\hat 0^+}\!=\!10\gev$ in $\mht\!-\!f\!/\!v$  plane, and the twin bottom quark mass $m_{\hat b}\!=\!15\gev$. The left and right panels of Fig.~\ref{fig:mthfovrmbt15mgb10} are for $\tilde\rho\!=\!0\gev$ and $\tilde\rho\!=\!0.5\gev$, respectively. For the chosen masses of twin hadrons we normalize average hadron multiplicity at the SM Higgs mass as $\langle N_{\hat G}\rangle_{125}\!=\!6$ and $\langle N_{[\hat b\bar{\hat b}]}\rangle_{125}\!=\!3$ for the twin glueball and bottomonium states, respectively. Note that for $m_{\hat b}\!=\!15\gev$ the twin bottomonium states with mass $m_{[\hat b\bar{\hat b}]}\simeq 30\gev$ are unstable and we assume they decay 100\% to the twin glueballs. Moreover, we set $\hat\kappa_0 \!=\!0.25$, the probability to produce the lightest glueball state $\hat G_{0^+}$. The shaded regions correspond to the unphysical parameters (gray), exclusion by the SM Higgs signal-strength measurements (red), and exclusion by the SM invisible branching fraction (hatched light-gray). Notice that heavy twin Higgs cross-section to twin glueballs $\hat G_{0^+}$ is comparable to its cross-section to the SM Higgs/vector bosons. The decay length of twin glueball $\hat G_{0^+}$ with mass $m_{\hat 0^+}\!=\!10\gev$ is rather large $\gsim \!10\,{\rm m}$ in most of the FTH parameter space, so it decays outside the detector and hence would appear as a missing energy signal.  

\begin{figure}[t]
\centering
\includegraphics[width=0.49\textwidth]{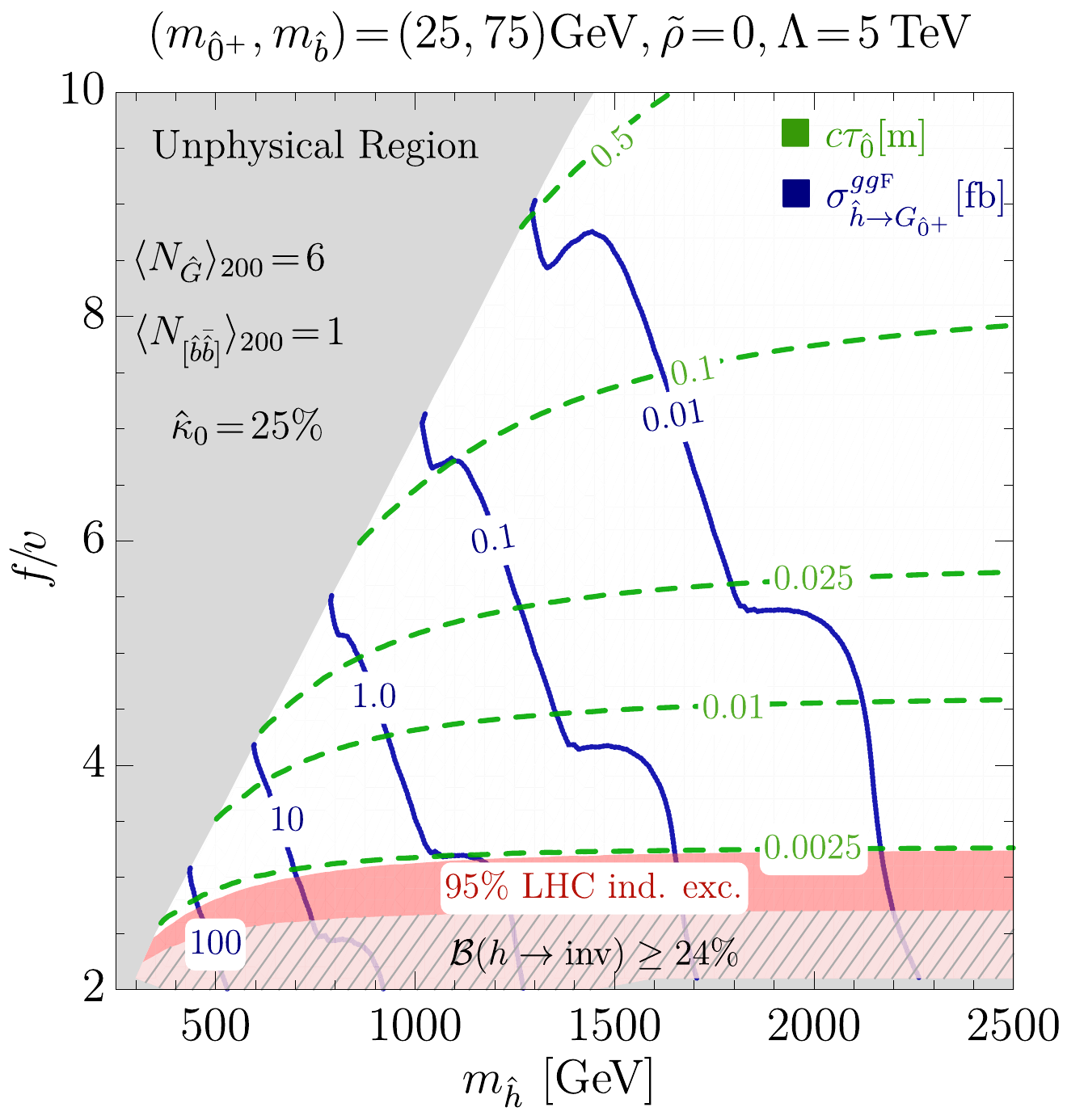}
\includegraphics[width=0.49\textwidth]{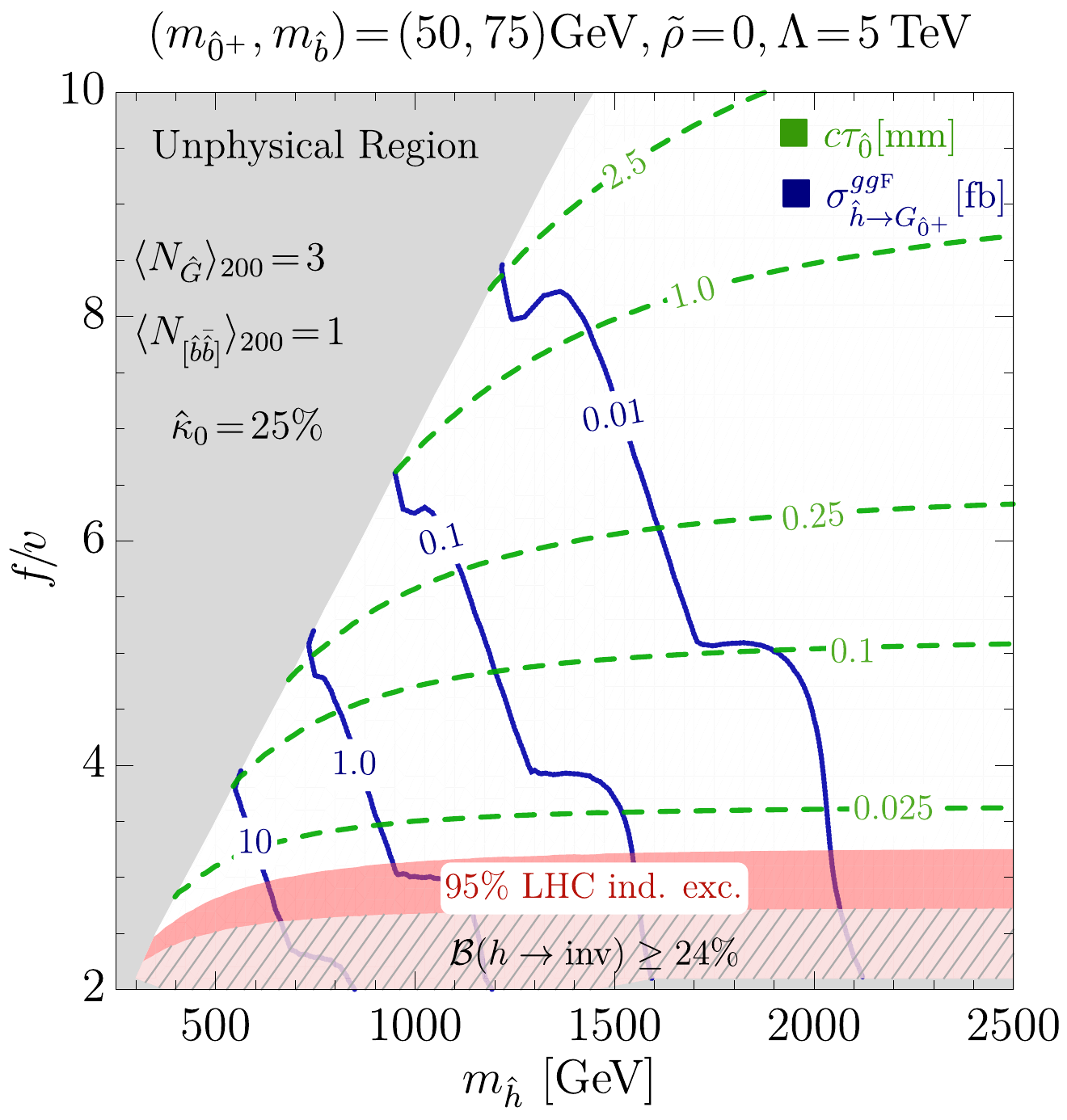}
\caption{These graphs show the contours of production cross-sections (solid blue) and proper decay-lengths (dashed green) of twin glueball $\hat G_{0^+}$ with mass $m_{\hat0^+}\!=\!25\gev$ and $50\gev$ in the left- and right-panel, respectively, with $m_{\hat b}\!=\!75\gev$ and $\tilde\rho\!=\!0$.}
\label{fig:mthfovr0mbt75mgb}
\end{figure}

For $0^{++}$ twin glueballs with masses $m_{\hat 0^+}\!=\!25\gev$ and $50\gev$ we show the contours of their production cross-sections (solid blue) and proper decay-lengths (dashed green) in $\mht\!-\!f\!/\!v$ plane in the left- and right-panel of Fig.~\ref{fig:mthfovr0mbt75mgb}, respectively. Whereas, in Fig.~\ref{fig:mthfovr0mbt75mgb}, we fix the twin bottom quark mass $m_{\hat b}\!=\!75\gev$, such that they annihilate to the twin glueballs. For the above masses of twin hadrons we normalized the average hadron multiplicity at $200\gev$ as $\langle N_{\hat G}\rangle_{200}\!=\!6$ for $m_{\hat 0^+}\!=\!25\gev$ and $\langle N_{\hat G}\rangle_{200}\!=\!3$ for $m_{\hat 0^+}\!=\!50\gev$, and $\langle N_{[\hat b\bar{\hat b}]}\rangle_{200}\!=\!1$. 
Notice for glueball mass $m_{\hat 0^+}\!=\!25\gev$ (left-panel of~\ref{fig:mthfovr0mbt75mgb}), its decay-length is in the range $10^{-2}\,{\rm m}\!\lsim \!c\tau_{0^{++}}\!\lsim\! 1\,{\rm m}$, which is an ideal range for the observation of displaced vertices at the LHC~\cite{ATLAS:2016olj,Aad:2015rba,Aad:2015uaa,CMS:2014wda,CMS:2014hka,Aaij:2016isa}. For glueball mass $m_{\hat 0^+}\!=\!50\gev$ (right-panel of~\ref{fig:mthfovr0mbt75mgb}), the decay-length is $10^{-5}\,{\rm m}\lsim\! c\tau_{0^{++}}\!\lsim\! 10^{-3}\,{\rm m}$, which would give prompt decays, however, as we have neglected any effects due to boosted production of the twin hadrons via heavy twin Higgs decays, which could potentially increase the decay-length. 

\begin{figure}[t]
\centering
\includegraphics[width=0.49\textwidth]{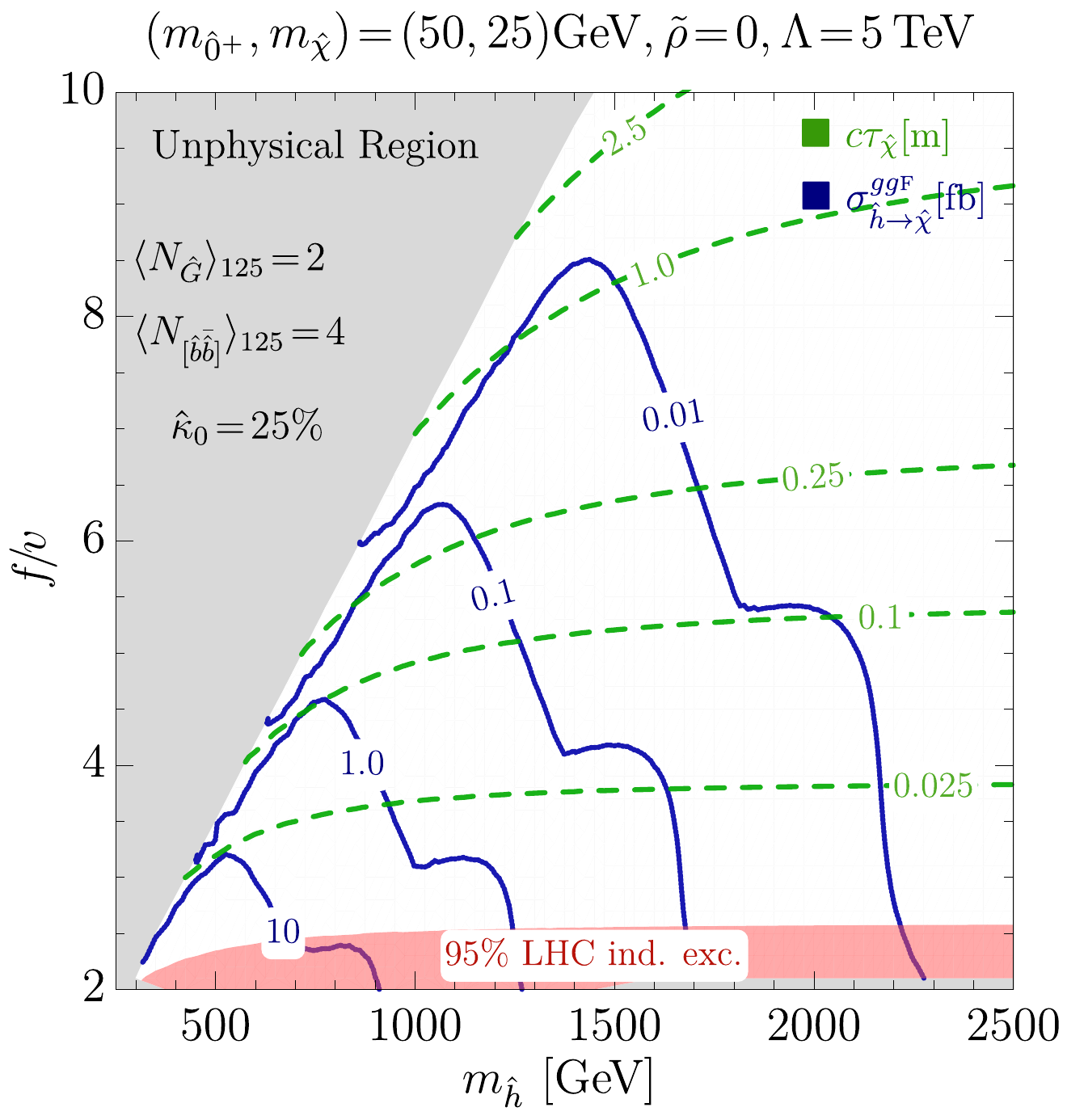}
\includegraphics[width=0.49\textwidth]{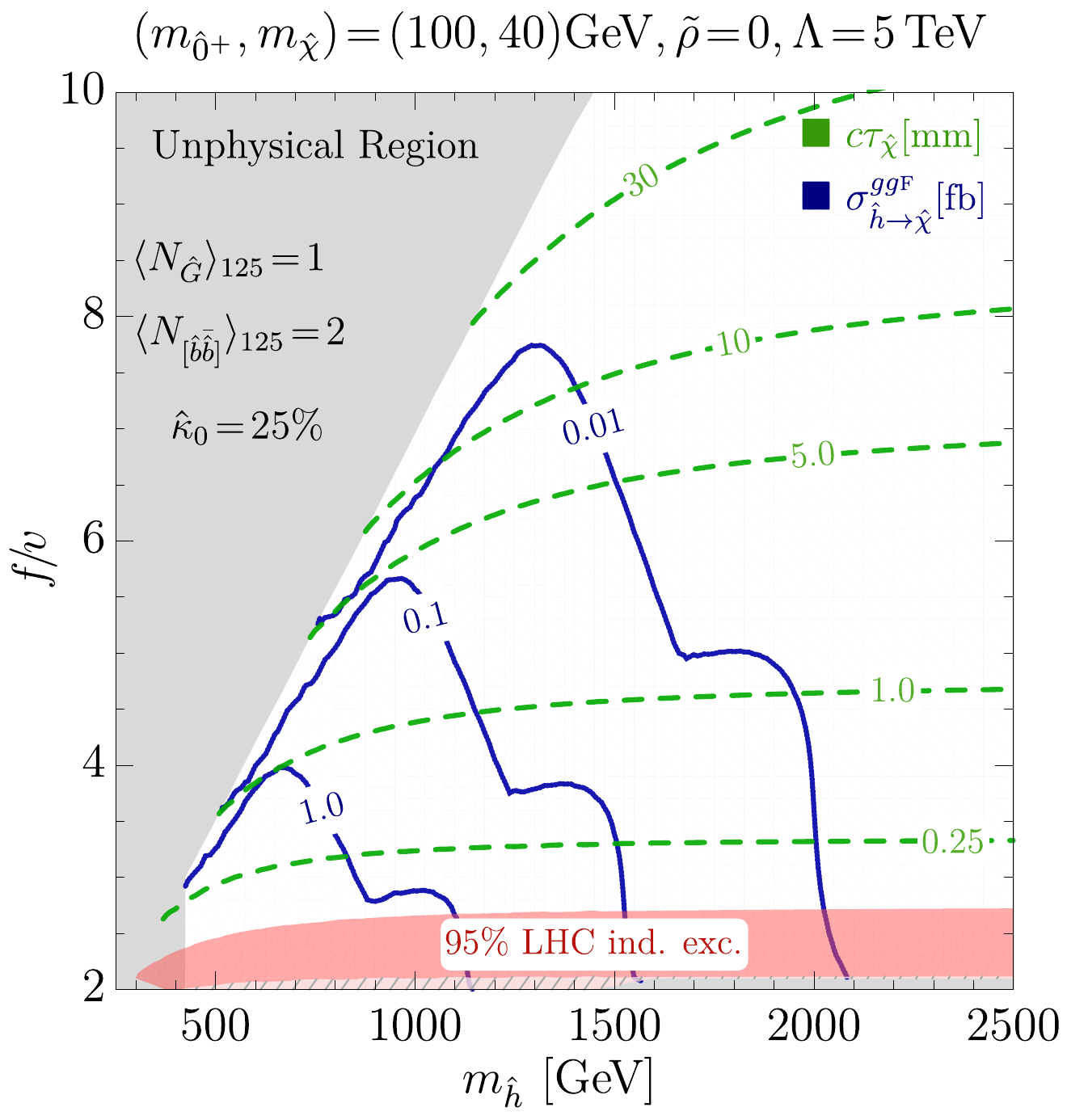}
\caption{These graphs show the contours of cross-sections (solid blue) and decay-lengths (dashed green) of twin bottomonium state $\hat \chi_{0}$ with mass $m_{\hat \chi}\!=\!25\gev$ and $40\gev$ in the left- and right-panel, respectively. Moreover, we consider $m_{\hat 0^+}\!=\!50\gev$ (left) and $100\gev$ (right) for $\tilde\rho\!=\!0$.}
\label{fig:mthfovr0mbtmgb}
\end{figure}

In Fig.~\ref{fig:mthfovr0mbtmgb}, we consider the case when the twin glueballs are heavier than the light bottomuium states such that they decay to two bottomonium states, \ie\ ${\rm glueballs\to [\hat b\bar{\hat b}][\hat b\bar{\hat b}]}$. In particular, we consider two cases with $(m_{\hat 0^+},m_{\hat \chi})\!=\!(50,25)\gev$ (left-panel) and $(m_{\hat 0^+},m_{\hat \chi})\!=\!(100,40)\gev$ (right-panel), where contours of the cross-section and the decay-length of twin bottomonium state $\hat\chi_0$ are shown. We normalize the average multiplicity at the SM Higgs mass as $(\langle N_{\hat G}\rangle_{125},\langle N_{[\hat b\bar{\hat b}]}\rangle_{125})\!=\!(2,4)$ and $(1,2)$ for the twin hadron masses $(m_{\hat 0^+},m_{\hat \chi})\!=\!(50,25)\gev$ and $(100,40)\gev$, respectively. Moreover, for the probability to produce $0^{++}$ bottomonium state $\hat \chi$ we take $\hat\kappa_0 \!=\!0.25$. Note that the twin bottomonium decay-length is in a range to give displaced vertices at the LHC. 

Current LHC searches~\cite{ATLAS:2016olj,Aad:2015rba,Aad:2015uaa,CMS:2014wda,CMS:2014hka,Aaij:2016isa} for long-lived particles are not sensitive for the cross-sections obtained here for the twin hadron production via heavy twin Higgs so we do not get any constraints, however in future some of the regions of the FTH parameter space can be probed by these searches. The studies of twin hadron production via the SM Higgs for the FTH model seems to offer more feasibility for the observation of twin hadrons at the LHC, see~\cite{Curtin:2015fna,Csaki:2015fba,Chacko:2015fbc,Pierce:2017taw}. Projected reach for the HL-LHC is to explore the TH parameter space up to $f\!/\!v\lsim 6$ in the twin glueball mass window $m_{\hat 0^+}\approx [10,60]$,~\cite{Curtin:2015fna}. However, the twin hadron production through the SM Higgs is limited only for the low mass ($\lsim\!m_h/2$), whereas, their production via heavy Higgs decay offers chances to probe heavier twin hadrons as well.

\subsection{Heavy Twin Higgs Phenomenology}
In this subsection, we briefly discuss the direct and indirect phenomenological prospects of the heavy twin Higgs in the weak sector. As we have learned from Fig.~\ref{fig:fthbrs}, the dominant decay modes of the twin Higgs in the FTH model are the SM and twin massive gauge bosons and the SM Higgs. Roughly speaking, phenomenology of heavy twin Higgs in the FTH model is very similar to that of the MTH case. The only difference, which can be crucial for indirect constraints, is the flexibility of twin light fermion masses which may enhance the SM Higgs invisible decay width. 

\begin{figure}[t]
\centering
\includegraphics[width=0.49\textwidth]{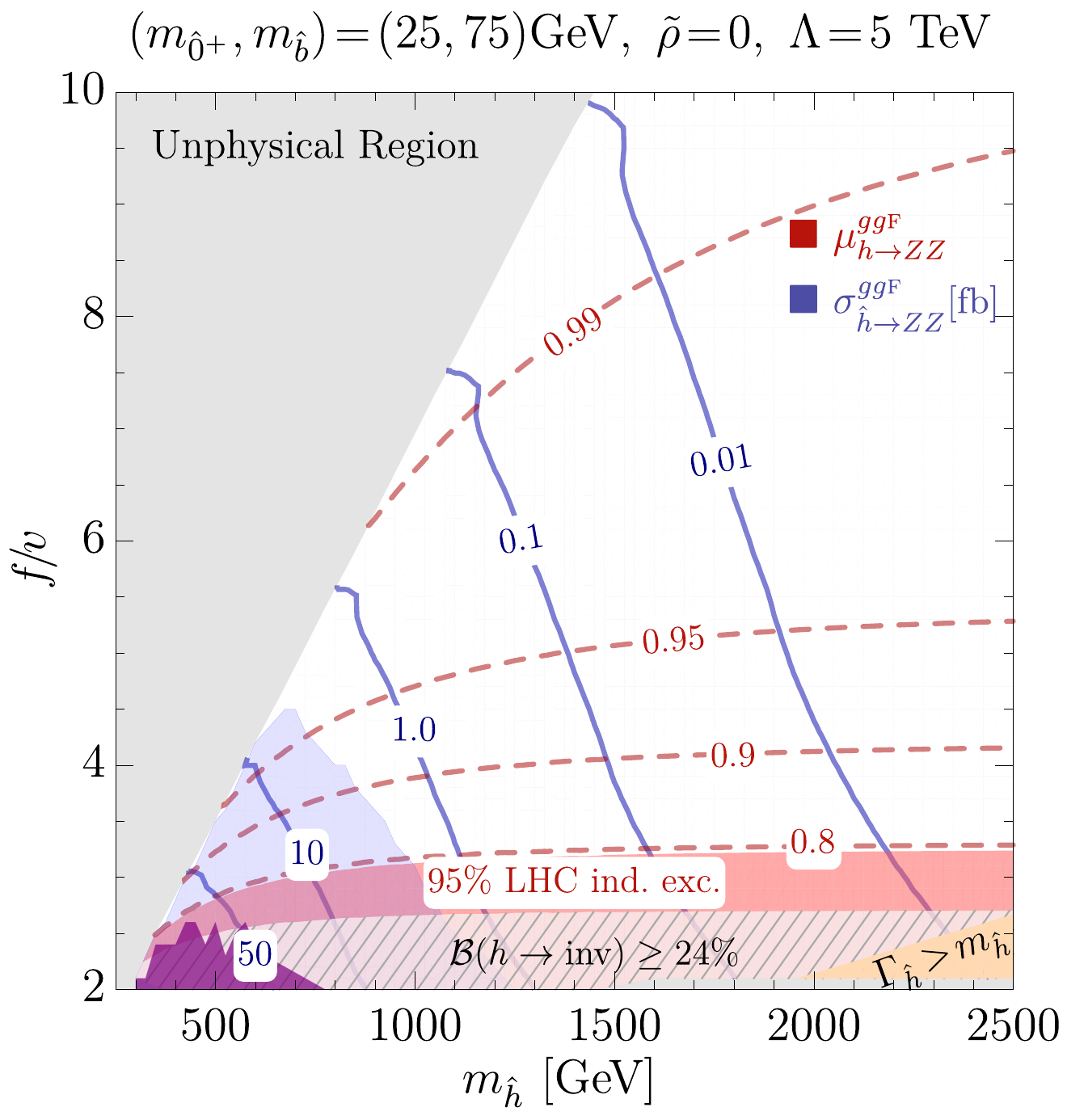}
\includegraphics[width=0.49\textwidth]{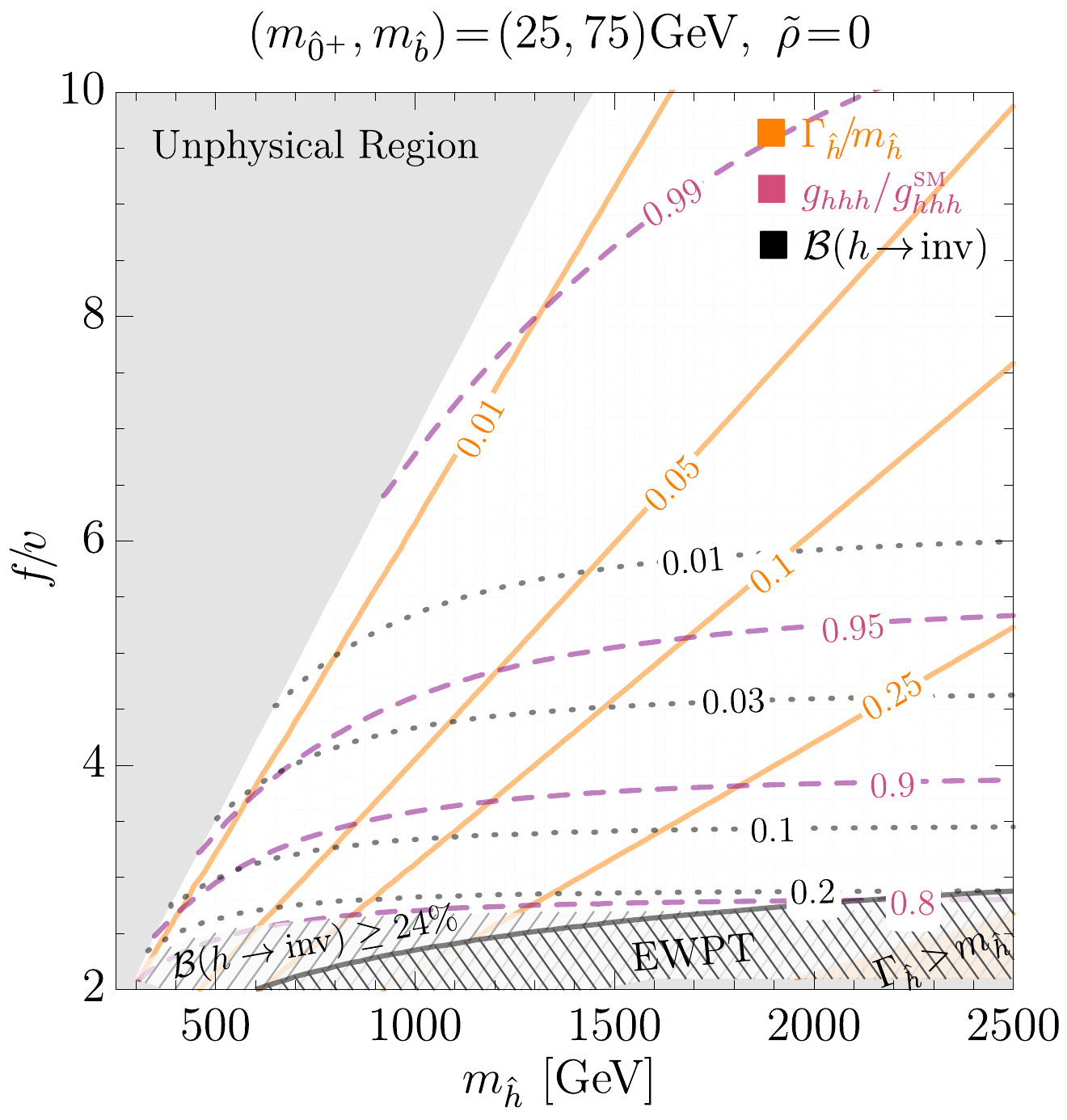}
\caption{The left-panel shows the contours of the cross-section of the heavy Higgs to the SM vector bosons (solid blue) and the SM Higgs signal-strength (dashed red) in $\mht\!-\!f\!/\!v$ plane of the FTH model with $\tilde \rho\!=\!0$. The right-plot show the contours of the ratio of heavy twin Higgs width over mass, $\Gamma_{\hat h}/\mht$ (solid orange), the ratio of the SM-like Higgs trilinear coupling over its corresponding SM values, $g_{hhh}/g^{\textsc{sm}}_{hhh}$ (dashed purple), and the SM Higgs invisible (twin) branching fraction ${\cal B}(h\!\to\! {\rm inv}.)$ (dotted). For these plots we take the twin glueball mass $m_{\hat 0^+}\!=\!25\gev$ and the twin bottom quark mass $m_{\hat b}\!=\!75\gev$, whereas the details of shaded regions are given in the text.}
\label{fig:mthzzinvwidthr0}
\end{figure}

We consider one benchmark scenario for the choice of twin glueball mass $m_{\hat 0^+}\!=\!25\gev$ and twin bottom mass $m_{\hat b}\!=\!75$. Moreover, we assume that twin lepton mass (Yukawa) scale as $m_{\hat \tau}\!=\!\tfrac{m_{\hat b}}{m_b}m_{\tau}$. In the left-panel of Fig.~\ref{fig:mthzzinvwidthr0} we plot contours of the twin Higgs cross-section (solid blue) to a pair of SM $Z$ bosons and the SM Higgs signal-strength $\mu^{gg\textsc f}_{h\!\to\!ZZ}$ (dashed red) for $\tilde\rho\!=\!0$ in $\mht\!-\!f\!/\!v$ plane. The right-panel of Fig.~\ref{fig:mthzzinvwidthr0} shows the concours of the ratio of heavy twin Higgs width over its mass $\Gamma_{\hat h}/\mht$ (solid orange), the SM Higgs trilinear coupling over its SM value $g_{hhh}/g^{\textsc{sm}}_{hhh}$ (shaded purple), and the SM Higgs invisible (twin sector) branching ratio ${\cal B}(h\!\to\! {\rm inv.})$ (dotted gray). The shaded regions represent: unphysical parameters (gray), exclusion by the SM Higgs signal-strength measurements (red), exclusion by the SM invisible branching fraction (hatched light-gray), exclusion by the EWPT at $95\%$ C.L (below the gray line, hatched area), exclusion by the direct heavy scalar searches at the LHC (purple), and the HL-LHC projected reach for direct searches (light blue). Note that for our choice of twin bottom mass $m_{\hat b}\!=\!75$, we have $m_{\hat \tau}\!\simeq\!30\gev$, which implies that the SM Higgs decays to $\hat b\bar{\hat b}$ are forbidden but to $\hat \tau\bar{\hat \tau}$ are allowed. The invisible decays of the SM Higgs are only in twin gluonic and twin leptonic channels.

\section{Conclusions} 
\label{Conclusions} 
In this work, we have explored phenomenological implications of the radial mode of Twin Higgs models, when the TH symmetry breaking is linearly realized. We consider an $SU(4)$ global symmetry for TH models which is spontaneously broken to $SU(3)$ and gives {\bf 7}~Goldstone bosons. Out of these {\bf 7}~Goldstone bosons, {\bf 6}~ are ``eaten'' by the SM and twin weak gauge bosons, and hence leaving only one Goldstone boson which plays the role of the SM Higgs boson. We consider an EFT approach and employ the most general symmetry breaking structure for the TH scalar potential including explicit soft and hard breaking $\mathbb{Z}_2$ discrete and $SU(4)$ global symmetries. These explicit breaking terms are the source of generating the SM Higgs mass. These terms are model dependent quantities and are calculable in a given perturbative UV complete TH model. The mixing of the SM-like Higgs and the twin Higgs (radial mode) provide a portal to the SM and twin sectors, which could potentially lead to the discovery of TH mechanism.

We have considered two Twin Higgs scenarios; the MTH, which has an exact mirror of the SM field content in the twin sector, and the FTH, which has the minimal field content needed to solve the little hierarchy problem. The symmetry breaking pattern of the TH mechanism is such that the SM Higgs and weak gauge bosons, and twin weak gauge bosons are the Goldstone bosons of same broken symmetry, therefore their interactions with the radial mode (heavy twin Higgs) are fixed. As a result, the heavy twin Higgs branching fractions to the SM Higgs and weak gauge bosons $W^\pm,Z$ are equal. Furthermore, the twin Higgs branching fractions to the SM sector as compared to the twin sector are approximately~$\sim\!4/3$. However, as shown in Secs.~\ref{Heavy Higgs Phenomenology in the Mirror Twin Higgs} and \ref{Heavy Higgs Phenomenology in the Fraternal Twin Higgs}, this universal behavior changes considerably when explicit hard breaking is turned on. Employing the current experimental direct and indirect searches at the LHC for a heavy scalar we explore the parameter space of the TH models. Current LHC direct searches for a heavy scalar to a pair of weak gauge boson constraint a small region of the MTH parameter space, in particular excluding the twin Higgs $\lsim500\gev$ mass. Whereas, the searches in the other decay channels especially to a pair of SM Higgs are not constraining with the available data. However, it is shown that the HL-LHC reach, at $14\tev$ with 3000~fb$^{-1}$, can discover/exclude the heavy twin Higgs up to $1\tev$ mass. On the other hand, indirect searches at the LHC by the SM Higgs signal-strength measurements are quite constraining. In particular, the current LHC data excludes $f\!/\!v\lsim 3$ without explicit hard breaking contributions. In the presence of explicit hard breaking contributions, the $f\!/\!v$ ratio could be less or more constrained depending on the sign of these terms. Moreover, the HL-LHC reach for the SM Higgs signal-strength measurements could potentially exclude the MTH parameter space up to $f\!/\!v\!\lsim\!6$.

In Sec.~\ref{Heavy Higgs Phenomenology in the Fraternal Twin Higgs}, we have explored phenomenology of the twin Higgs in the FTH model. In particular, we have discussed the consequences of twin QCD confinement. In the absence of light quark flavors in the FTH model, twin QCD confinement/hadronization leads to the formation of twin glueball/bottomonium states, which have interesting phenomenological signatures. We have discussed the production of these twin hadrons via heavy twin Higgs decays at the LHC. More importantly, the $0^{++}$ glueball $\hat G_{0^+}$ and bottomonium $\hat \chi_0$ states have the correct quantum number to mix with the scalar sector of the model, as a result, decays of these twin hadrons back to the SM light fermions are possible. The lessons we have learned from the twin hadron phenomenology are: (i)~The heavy Higgs cross-section to the $0^{++}$ twin hadrons at the LHC is comparable to that of the SM vector bosons, and (ii)~Depending on the masses of $0^{++}$ twin hadrons, their decays can be prompt, displaced or even can escape the detectors. These signatures offer a unique experimental opportunity to search for multi-jet prompt or displaced objects and/or missing energy signals at the colliders. Hence, any signals of such events at the LHC could lead to the discovery of twin Higgs framework. 

\subsection*{Acknowledgements}
The author thanks Zackaria Chacko and Saereh Najjari for discussions and comments on the draft, and also informing about their related work~\cite{Chacko:2017xpd}. He also thanks Barry Dillon and Bohdan Grzadkowski for useful comments. He is grateful to the Mainz Institute for Theoretical Physics (MITP) and the Laboratoire de Physique Th\'eorique (LPT) Orsay, for their hospitality and support while this work was in progress. This work was partially supported by the National Science Centre (Poland) research projects, decision DEC-2014/13/B/ST2/03969, and DEC-2014/15/B/ST2/00108.

\appendix
\section{Twin Higgs Feynman Rules and Partial Decay Widths}
\label{Partial decay widths of heavy twin Higgs}
In this Appendix, we collect partial decay widths of the heavy twin Higgs employed in the main text. The most relevant Feynman rules for the scalar sector of the TH models are collected in Fig.~\ref{fig:couplings}, where the triangle loop functions are given below. 
\begin{figure} [t]
\centering
\includegraphics[width=\textwidth]{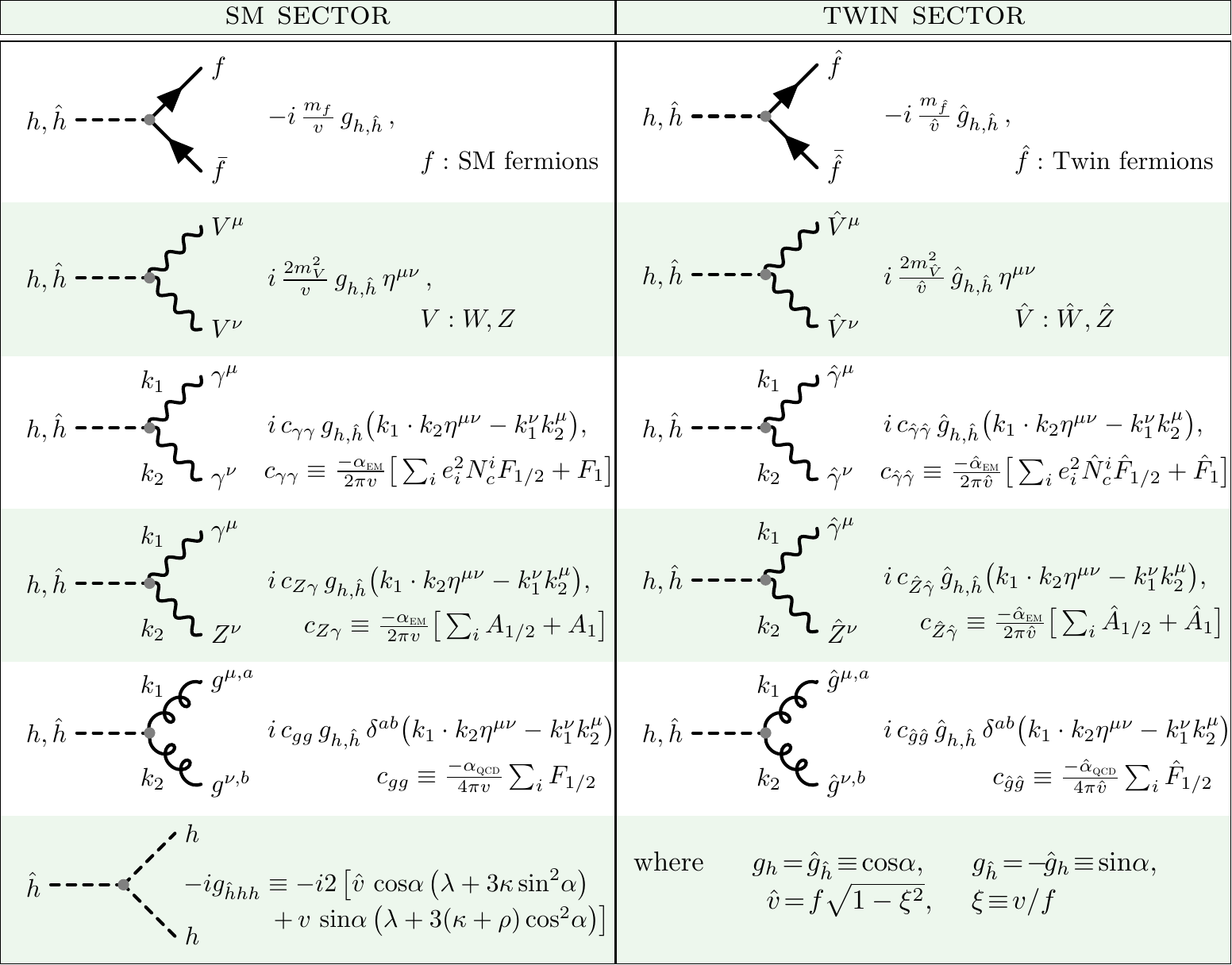}
\caption{Feynman rules for the interactions of SM and twin sector particles with the SM Higgs $h$ and heavy twin Higgs $\hat h$. The SM triangle loop functions $F_{1/2}$, $F_1$, $A_F$ and $A_W$ are given below and the corresponding twin triangle loop functions $\hat F_{1/2}$, $\hat F_1$, $\hat A_F$ and $\hat A_W$ have similar form as the SM ones but with twin particles in the loops.}
\label{fig:couplings}
\end{figure}

\paragraph{Heavy Higgs decays into fermions:} The two body decays of heavy twin Higgs into the SM and twin sector fermions are given as:
\beq
\Gamma^{\hat h}(f f)=\frac{G_{\textsc f}g_{\hat h}^2}{4\sqrt2\pi}\mht m_f^2\bigg(1-\frac{4m_f^2}{\mht^2}\bigg)^{3/2}, \lsp
\Gamma^{\hat h}(\hat f \hat f)=\frac{\hat G_{\textsc f}\hat g_{\hat h}^2}{4\sqrt2\pi}\mht m_{\hat f}^2\bigg(1-\frac{4m_{\hat f}^2}{\mht^2}\bigg)^{3/2},   \label{drate_hff}
\eeq
where $g_{\hat h}\!=\!\sin\alpha$, $\hat g_{\hat h}\!=\!\cos\!\alpha$, and twin Fermi constant is $\hat G_{\textsc f}\!=\!(v^2/\hat v^2)G_{\textsc f}$. Moreover, in our numerical analysis we have allowed off-shell decays of the heavy fermions (the SM and the twin top quarks).

\paragraph{Heavy Higgs decays into gauge bosons:} The heavy twin Higgs partial decay width to on-shell SM massive gauge bosons is, 
\begin{align}
\Gamma^{\hat h}(V_1V_2)&=\frac{G_{\textsc f}\,\mht^3\, g_{\hat h}^2}{8\sqrt2\,\pi\,{\cal S}_V} \lambda\big(x_1,x_2\big) \Big[\lambda^2\big(x_1,x_1\big)
    +12 \,x_1\,x_2\Big] 
\end{align}
where, $\lambda(x_1,x_2)\equiv\sqrt{(1-x_1-x_2)^2-4x_1 x_2}$ with $x_{1,2}=m_{V_{1,2}}^2/\mht^2$ and ${\cal S}_V$ is a symmetry factor, for identical gauge bosons it is 2 otherwise 1. Moreover, we also include the heavy Higgs decays with off-shell massive gauge bosons, which allows four-body decays of the heavy Higgs, as:
\beq
 \Gamma ^{\hat h}(V^\ast_1 V^\ast_2)\!=\!\frac{1}{\pi^2}\!\!\int_0^{\mht^2} \!\!\!\frac{ds_1\; m_{V_1}\Gamma_{V_1}}{\big(s_1-m_{V_1}^2\big)^2+m_{V_1}^2\Gamma_{V_1}^2} \int_0^{\left(\mht-\sqrt{s_1}\right)^2} \hspace{-17pt}\frac{ds_2 \;m_{V_2}\Gamma_{V_2}}{\big(s_2 -m_{V_2}^2\big)^2+m_{V_2}^2\Gamma_{V_2}^2}\
 \Gamma^{\hat h} (V_1V_2)\, ,
\eeq
where  $s_{1,2}$ are the squared invariant masses of the intermediate gauge bosons and $\Gamma_V$ being their total decay widths. 

The partial decay widths of the twin Higgs to massless gauge bosons of the SM are,
\begin{align}
\Gamma^{\hat h}(gg)&=\frac{G_{\textsc f}\alpha_{\textsc{qcd}}^2\mht^3}{64\sqrt2\pi^3}\Big\vert\sum_{i=t,b} F_{1/2}(\tau_i)\Big\vert^2 \; g_{\hat h}^2,	\\
\Gamma^{\hat h}(\gamma\gamma)&=\frac{G_{\textsc f}\alpha_{\textsc{em}}^2\mht^3}{128\sqrt2\pi^3}\Big\vert\sum_{i=t,b}e_i^2N_c^i  F_{1/2}(\tau_i)+F_{1}(\tau_W)\Big\vert^2\;g_{\hat h}^2,\\
\Gamma^{\hat h}(Z\gamma)&=\frac{G_{\textsc f}\alpha_{\textsc{em}}\mht m_Z^2}{64\pi^4}\swsq\cwsq\bigg(1-\frac{m_Z^2}{\mht^2}\bigg)^3\Big\vert \sum_{i=t,b}A_{1/2}+A_1\Big\vert^2 g_{\hat h}^2,
\end{align}
where the triangle loop functions $F_{1/2}$, $F_1$, $A_F$ and $A_W$ are adopted from the Higgs Hunter's Guide~\cite{Gunion:1989we} and for completeness we collect them below. 

It is straightforward to obtain the heavy twin Higgs partial decay widths to the twin gauge bosons from the above relations by interchanging $V\to \hat V,\, g_{\hat h}\to \hat g_{\hat h},\, G_{\textsc f}\to \hat G_{\textsc f},\, \alpha_{\textsc{em}}\to \hat \alpha_{\textsc{em}},\, \alpha_{\textsc{qcd}}\to\hat  \alpha_{\textsc{qcd}}\, F_{1/2}\to \hat F_{1/2}, \,F_1\to \hat F_1,\, A_F\to \hat A_F$ and $A_W\to \hat A_W$.

\paragraph{Heavy Higgs decays into SM di-Higgs:}
The heavy twin Higgs partial decay width to a pair of the SM Higgs is,
\beq
\Gamma^{\hat h}(hh)=\frac{g_{\hat hhh}^2}{32\pi}\frac1\mht \sqrt{1-\frac{4m_h^2}{\mht^2}}~, 
\eeq
where the trilinear coupling $g_{\hat hhh}$ is given in \eqref{gthhh}.

\paragraph{The loop functions:} Following the notations of the Higgs Hunter's Guide~\cite{Gunion:1989we}, the form factors $F_{1/2}$, $F_1$, $A_F$ and $A_W$ are
\begin{align}
F_{1/2}(\tau_i)&=-2\tau_i\big[1+(1-\tau_i)f(\tau_i)\big], 
\lsp F_{1}(\tau_i)=2+3\tau_i+3\tau(2-\tau_i)f(\tau_i), \\
A_{1/2}(\tau_i,\kappa_i)&=\frac{-e_iN_c^i}{\sw\cw}\big(1-4e_i\swsq\big) \big[{\cal I}_1(\tau_i,\kappa_i)-{\cal I}_2(\tau_i,\kappa_i)\big]		,\\
A_1(\tau_i,\kappa_i)&=-\frac{\cw}{\sw}\Big[4\Big(3-\frac{\swsq}{\cwsq}\Big){\cal I}_2(\tau_W,\kappa_W)	\notag\\
&\hspace{2cm} + \Big((1+2/\tau_W)\frac{\swsq}{\cwsq}- (5+2/\tau_W)\Big){\cal I}_1(\tau_W,\kappa_W)\Big],		
\end{align}
where $\tau_i\equiv4m_i^2/\mht^2$, $\kappa_i\equiv 4m_i^2/m_Z^2$, and 
\begin{align}
{\cal I}_1(\tau_i,\kappa_i) &= \frac{\tau_i\kappa_i}{2(\tau_i-\kappa_i)}+ \frac{\tau_i^2\kappa_i^2}{2(\tau_i-\kappa_i)^2}\Big(f(\tau_i)-f(\kappa_i)\Big)+ \frac{\tau_i^2\kappa_i}{(\tau_i-\kappa_i)^2}\Big(g(\tau_i)-g(\kappa_i)\Big),	\\
{\cal I}_2(\tau_i,\kappa_i) &= -\frac{\tau_i\kappa_i}{2(\tau_i-\kappa_i)}\Big(f(\tau_i)-f(\kappa_i)\Big),	\\
f(\tau_i)&= \begin{cases}
\arcsin^2\big(1/\sqrt{\tau_i}\big), & \hspace{0.7cm}\text{if }  \tau_i \geq 1,\\    
-\frac14\big[\ln\big(\frac{1+\sqrt{1-\tau_i}}{1-\sqrt{1-\tau_i}}\big)-i\pi\big]^2, & \hspace{0.7cm}\text{if }  \tau_i <1,
\end{cases}\\
g(\tau_i)&= \begin{cases}
\sqrt{\tau_i-1}\arcsin^2\big(1/\sqrt{\tau_i}\big), & \text{if }  \tau_i \geq 1,\\    
-\frac{\sqrt{1-\tau_i}}2\big[\ln\big(\frac{1+\sqrt{1-\tau_i}}{1-\sqrt{1-\tau_i}}\big)-i\pi\big]^2, & \text{if }  \tau_i <1.
\end{cases}
\end{align}
Note that the corresponding twin sector loop functions are straightforward to obtain from above generic formulae by appropriately replacing the masses, couplings, etc.

\providecommand{\href}[2]{#2}\begingroup\raggedright\endgroup


\end{document}